\documentclass[useAMS,usenatbib]{mn2e}
\usepackage[dvips]{graphicx}
\usepackage{amssymb,amsmath}
\usepackage{fleqn,subfigure,lscape,epsfig,graphics}
\usepackage{times}
\usepackage[dvipsnames,usenames]{color}
\usepackage{colortbl}

\title[Measuring gravitational lens time delays]{Measuring gravitational lens time delays using low-resolution radio monitoring observations}
\author[G\"urkan et al.]
{
{\parbox{\textwidth}{G. G\"urkan$^{1,2}$\thanks{E-mail:g.gurkan-uygun@herts.ac.uk},
N. Jackson$^{1}$,
L.V.E. Koopmans$^{3}$,
C.D. Fassnacht$^{4}$, 
A. Berciano Alba$^{5}$ \vspace{0.4cm}\\}
}\\
$^{1}$Jodrell Bank Centre for Astrophysics, Alan Turing Building, University of Manchester, Manchester M13 9PL, U.K.\\
$^{2}$ School of Physics, Astronomy and Mathematics, University of Hertfordshire, College Lane, Hatfield AL10 9AB, U.K.\\
 $^{3}$ Kapteyn Astronomical Institute, University of Groningen, P.O. Box 800,
9700 AV Groningen, The Netherlands\\
$^{4}$ Department of Physics, University of California, Davis, 1 Shields Avenue,
Davis, CA 95616, USA\\
$^{5}$ Netherlands Foundation for Research in Astronomy (ASTRON), Postbus 2, 7990 AA Dwingeloo, The Netherlands\\
}

\begin{document}

\def\aj{AJ}					
\def\araa{ARA\&A}				
\def\apj{ApJ}					
\def\apjl{ApJL}					
\def\apjs{ApJS}					
\def\apss{Astrophysics and Space Science}
\def\capsp{Comments on Astrophysics and Space Physics}
\def\aap{A\&A}					
\def\aapr{A\&A~Rev.}				
\def\aaps{A\&AS}				
\def\azh{AZh}					
\def\baas{BAAS}					
\def\jrasc{JRASC}				
\def\memras{MmRAS}				
\def\mnras{MNRAS}					
\def\pasp{PASP}					
\def\pasj{PASJ}					
\def\qjras{QJRAS}				
\def\skytel{S\&T}				
\def\solphys{Sol.~Phys.}			
\def\sovast{Soviet~Ast.}			
\def\ssr{Space~Sci.~Rev.}			
\def\zap{ZAp}					
\def\na{New Astronomy}				
\def\iaucirc{IAU~Circ.}				
\def\aplett{Astrophys.~Lett.}			
\def\apspr{Astrophys.~Space~Phys.~Res.}		
\def\bain{Bull.~Astron.~Inst.~Netherlands}	
\def\memsai{Mem.~Soc.~Astron.~Italiana}		

\def\ao{Appl.~Opt.}				

\def\pra{Phys.~Rev.~A}				
\def\prb{Phys.~Rev.~B}				
\def\prc{Phys.~Rev.~C}				
\def\prd{Phys.~Rev.~D}				
\def\pre{Phys.~Rev.~E}				
\def\prl{Phys.~Rev.~Lett.}			
\def\nat{Nature}				
\def\fcp{Fund.~Cosmic~Phys.}			
\def\gca{Geochim.~Cosmochim.~Acta}		
\def\grl{Geophys.~Res.~Lett.}			
\def\jcp{J.~Chem.~Phys.}			
\def\jgr{J.~Geophys.~Res.}			
\def\jqsrt{J.~Quant.~Spec.~Radiat.~Transf.}	
\def\nphysa{Nucl.~Phys.~A}			
\def\physrep{Phys.~Rep.}			
\def\physscr{Phys.~Scr}				
\def\planss{Planet.~Space~Sci.}			
\def\procspie{Proc.~SPIE}			
\def\rpp{Rep.~Prog.~Phys.}			
\let\astap=\aap
\let\apjlett=\apjl
\let\apjsupp=\apjs
\let\applopt=\ao
\let\prep=\physrep


\date{Accepted ...... Received ...... ; in original form......   }

\pagerange{\pageref{firstpage}--\pageref{lastpage}} \pubyear{2012}
\maketitle
\label{firstpage}
\begin{abstract}

Obtaining lensing time delay measurements requires long-term monitoring campaigns with a high enough resolution ($<1$$^{\prime\prime}$) to separate the multiple images. In the radio, a limited number of high-resolution interferometer arrays make these observations difficult to schedule. To overcome this problem, we propose a technique for measuring gravitational time delays which relies on monitoring the total flux density with low-resolution but high-sensitivity radio telescopes to follow the variation of the brighter image. This is then used to trigger high-resolution observations in optimal numbers which then reveal the variation in the fainter image. We present simulations to assess the efficiency of this method together with a pilot project observing radio lens systems with the Westerbork Synthesis Radio Telescope (WSRT) to trigger Very Large Array (VLA) observations. This new method is promising for measuring time delays because it uses relatively small amounts of time on high-resolution telescopes. This will be important because instruments that have high sensitivity but limited resolution, together with an optimum usage of followup high-resolution observations from appropriate radio telescopes may in the future be useful for gravitational lensing time delay measurements by means of this new method.

\end{abstract}


\section{Introduction}
 
The strong gravitational lensing effect occurs when light from a 
background source (a galaxy or a quasar) is deflected by the gravitational 
field of an intervening mass, such as a galaxy or cluster of galaxies,
forming multiple images of the background source \citep{1992book1,2006book2}. This phenomenon
is widely used in astrophysics and cosmology as a tool because it
provides information about mass distributions in the lensing object \citep [e.g.][]{1991new36,2002new37,2007new42} as well as magnified views of the sources \citep [e.g.][]{2007new9,2011ref00}.

\cite{1964ref3} demonstrated that lensing time
delays can be used to measure cosmological distances, in particular the Hubble
constant $H_{0}$. This can be done if the background source is variable by
measuring time delays between variations of the images, thereby deducing an 
absolute distance scale provided the redshifts of the source and lens, and 
the mass model of the lens potential, can be determined. The time delay 
in a lens system scales with the size of the Universe and inversely
with $H_{0}$; in a given system, it also depends
on other cosmological parameters such as the matter density $\Omega_{m}$ and
dark energy density $\Omega_{\Lambda}$, although this dependence is relatively weak. Consequently, large-scale time delay studies
in future may allow these parameters to be determined as well
\citep{2009new29,2010time13,2013ref77}. It is worthwhile to note that these parameters affect the
$H_0$ determination at a relatively low level, and in principle gravitational
lensing is therefore a useful one-step method for $H_0$ determination on
cosmological scales. A number of groups are 
currently carrying out monitoring campaigns to determine time delays
for lenses in the optical
\citep[e.g.][]{2005new22,2006time34,2007new23,2008new27,2008new24,2011new25,2013ref76,2013ref79}. Measured time delays by means of these projects generally suggest $63<H_{0}<82$ km s$^{-1}$Mpc$^{-1}$. See e.g. \cite{2007ref7} and \cite{2010ref64}
for more general reviews of measurements of the Hubble constant.

A difficulty with obtaining lensing time delay measurements is that it requires monitoring campaigns of months to years with a high enough resolution ($<1$$^{\prime\prime}$) to separate the multiple images. The four-image lens 
system B1608+656 \citep{1995ref58,1995ref59}, for instance, required observations for multiple seasons with the VLA. After almost 3 years' monitoring of B1608+656, the accuracy of the time delays improved by 
factors of 2-3 due to an increase of the flux density of the background source by 25$\%$ \citep{1999time3,2002time4}.

To minimise the problems mentioned above, we propose a new method for
gravitational lens time delay measurements. In asymmetric double image and long-axis quadruple image lens systems we can take advantage of the fact that the brighter image(s) varies first and dominates the total flux. This method builds on a suggestion by \cite{1996method1} who proposed using low-resolution observations only. Low-resolution but high sensitivity observations are used which are sufficient to recognise the variation of the brighter image. Afterwards, observations with a high-resolution interferometer array are triggered to see the variation of the fainter images. In order to assess the efficiency of our technique we performed cross-correlation
simulations using the Pelt dispersion statistic \citep{1996method9} and artificial light curves. We also used the Pelt dispersion statistic to evaluate the results of our pilot project.

This paper is organised as follows. A description of our proposed technique, together with results from simulations performed to assess its efficiency, are presented in Section 2. As a pilot project, a flux monitoring
campaign was carried out with the WSRT at 5 GHz including 39 epochs of
observations. VLA observations at 5 GHz giving 1$^{\prime\prime}$ 
resolution were triggered to resolve the images of the system B1030+074 
which showed a possible variability feature during the flux monitoring. 
These results are shown in section 3 and 4. Finally, in section 5 we 
discuss this technique and the results.
\begin{table*}
\begin{tabular}{cccccccc}
\hline
\hline
Object&Type&Separation&Flux&Flux&Likely&Phase&References\\
&&(arc-sec)&Brighter&Fainter&delay&Calibrators&\\
&&&Image&Image&&\\
&&&(mJy)&(mJy)&(days)&\\
\hline
\hline
CLASS B0445+123&D&1.2&25&4&30&3C138&\cite{2003ref1}\\
CLASS B0631+519&D&1.2&34&5&15&3C147&\cite{2005ref2}\\
CLASS B0850+054&D&0.7&55&9&18&J0907+037&\cite{2003ref4}\\
CLASS B0739+366&D&0.6&27&5&10&J0736+331&\cite{2001ref3}\\
JVAS B1030+074&D&1.6&200&13&110&J1015+089&\cite{1998ref5}\\
CLASS B1152+199&D&1.6&50&18&30&J1142+185&\cite{1999ref6}\\
JVAS B1422+231&Q&1.2&500&5&25&J1429+218&\cite{1992ref7}\\
CLASS B2319+051&D&1.4&56&11&25&J2398+034&\cite{2001ref8}\\
\hline
\end{tabular}
\caption[Features of the lens systems.]{The table shows the features of the
  target lenses. The lenses are selected among double or long-axis quadruple CLASS
 lenses with the highest flux ratios. D and Q refer to $\textit{Double lenses (2-image lenses)}$
  and $\textit{Quadruple lenses (4-image lenses)}$, respectively. Separation
  between the images of the sources are given in column 3. Column 4
  gives the flux density of the component which varies first. Column 5 gives
  the flux density of the delayed component and column 6 the time delay if
  $H_{0}$ = 70 km s$^{-1}$ Mpc$^{-1}$ (assuming an isothermal mass profile for the lens galaxy). The calibrator
sources used during the total flux monitoring can be seen in column 7.} \label{lens}
\end{table*}

\section{A Method for Time Delay Measurements}

 There are only $\sim$20 gravitational lens systems  which have time delay
 measurements among which 5 lens systems have radio light curves and 17 lenses
 have optical light curves. The main reason for this is that there are fewer radio
 lenses that show significant variation. It should be also noted that lensing time delay
 measurements require long-time monitoring campaigns with a high-resolution ($<1^{\prime\prime}$)
 telescope to separate the images of a lensed source. In
 the radio, the VLA, MERLIN (Multi-Element Radio Linked Interferometer Network), VLBA and LOFAR (Low-frequency Array) \citep{2013ref85} are the only interferometer arrays that are capable of regular imaging
 with the required resolution \citep{2002time4,2001method18}. 

\cite{1996method1} proposed a ``light curve reconstruction'' method for the 
determination of time delays in gravitational lens systems. They suggested
that it is possible to reconstruct the light curves of the individual images
using a single dish total flux monitoring by assuming values for the time
delay and the magnification ratio. However, the evaluation of the effects of different parameters on the method (e.g. magnification ratio, observing period) showed that additional interferometric observations are necessary in order to achieve significant results. For this reason, they concluded that a few additional interferometric observations are necessary. The true time delay value can then be
determined by checking the consistency of the reconstructed light curves
utilising the flux density ratio of the images by way of additional interferometric observations.


Here we propose a technique which has similar features to the method used by
\cite{1996method1}. This technique which proposes using observations of both low and
high-resolution radio interferometer arrays, is observationally more
complicated, but minimises the required time of high-resolution observations. We focus on observations at radio frequencies as the radio fluxes of lensed
images are not affected by micro-lensing produced by stars in the
lens galaxy \citep[e.g.][]{1979ref90,1989ref89,1990ref91}, the presence of dust \citep[e.g.][]{2006ref88}, or the confusion between the lens galaxy
and the lensed images. We note that extrinsic effects can affect radio light curves of objects that are close to the galaxy disk \citep{2000ref86,2003ref0}. By contrast, optical lenses may suffer from all
these problems. Among radio lenses, asymmetric double image and long-axis
quadruple image lenses are particularly interesting because the
brighter image shows the intrinsic variation first on the lensed source
flux and after a time the fainter component(s) varies. In the case of long-axis quadruples three close images act
like a brighter image and the time delay between three components is
  much smaller than the delay between the faint image and the bright component. Undertaking total flux
monitoring gives us the 
variation of the brighter component, which dominates the total flux. It is therefore 
possible to use low-resolution but highly sensitive radio observations for total 
flux monitoring. Once a
light curve shows signs of variation in the total flux of the lens, high
resolution observations can be triggered at the time that the fainter component
is expected to vary. Then, a reasonable period of followup using the high-resolution monitoring should allow us to recover a time delay assuming that $H_{0}$ is close to 70 km s$^{-1}$ Mpc$^{-1}$ ($\sim$20-50$\%$) and that a reasonable model for the lens galaxy's mass profile is in hand \citep{2006method17,2009referans49,2010method20} (although there is a degeneracy that couples the mass density profile and time delays and this affects the derived $H_{0}$). At present, monitoring telescopes need to be in the northern hemisphere because of the availability of the VLA or MERLIN for followup, but this will change in the future with the advent of the SKA.

\subsection{The Pelt Dispersion Statistic}
For light curves of two images, A and B, the Pelt statistic \citep{1994method7,1996method8,1996method9} is calculated by delaying one light curve by $\tau$ with respect to the other and measuring the dispersion of the difference between the two, using a variable scaling factor $\mu$. The value of $\tau$ for which the statistic is a minimum is the presumed time delay. Suppose we have a dataset ($t_i$, $t_j$) of individual
images; a brighter, $A_i$, and a fainter, $B_j$. When the composite light curve, $C_k$, is generated, the fluxes of $B_j$ are multiplied by a scaling factor $\mu$ 
and the data points of $B_j$ are shifted by a delay $\tau$:

\begin{equation}
C_k(t_k)= 
\begin{cases} A_i, & \text{if}\quad{}t_k = t_i
\\
\mu{}B_j, &\text{if}\quad{}t_k = t_j+\tau
\end{cases}
\end{equation}

and the dispersion $D^2$ of the scatter around the composite light curve is
estimated:

\begin{equation}
D^2=\underset{\mu}{\operatorname{min}}\frac{\sum_{k=1}^{K-1}W_{k,k+1}G_{k}(C_{k+1}-C_{k})^2}{2\sum_{k=1}^{K-1}W_{k,k+1}G_{k}}
\end{equation}

where $G_{k}$=1 only when $C_{k+1}$ and $C_{k}$ are from different images and 
$G_{k}$=0 otherwise. The accuracy of the observations is taken into account by
using the statistical weights ($W_{i}$  and $W_{j}$) of the combined light curve data:

\begin{equation}
W_{k}=W_{i,j}=\frac{\mu W_{i}W_{j}}{\mu W_{i}+W_{j}}
\end{equation}
where $\textit{k}$ = 1,...,$\textit{N}$.

For the technique presented here, the detectability of a time delay depends on
a large number of parameters. These can be divided into parameters associated
with the source, namely the flux ratio of the lensed images and the difference between the
brightest and faintest part of the light curve in any definite feature (the
amplitude of variation), and
those associated with the observations: the timing of the sequence of
triggered observations, the time interval between two consecutive flux
measurements (the sampling frequency), number of epochs, errors in the flux measurement and the noise level of both the low-resolution and high-resolution triggered observations.

\subsection{Light curve simulations}

There are two potential problems with the monitoring approach we propose here.
The first is that the total light curve contains flux from both components,
thus affecting the measured time delay because the delay analysis effectively
compares the total flux density with the fainter image flux density, rather
than the brighter image with the fainter. This problem becomes worse for
a lower flux ratio, for which the contamination of the total light curve
is worse, and for shorter time delays, for which the timescale of variability
and the time delay may be similar. The second problem is that a triggering
strategy must be chosen which is optimally adjusted, in number and separation
of samples, to achieve the best result for the time delay without using
large numbers of triggered observations.

\begin{figure*}
\begin{tabular}{c}
\includegraphics[width=9cm]{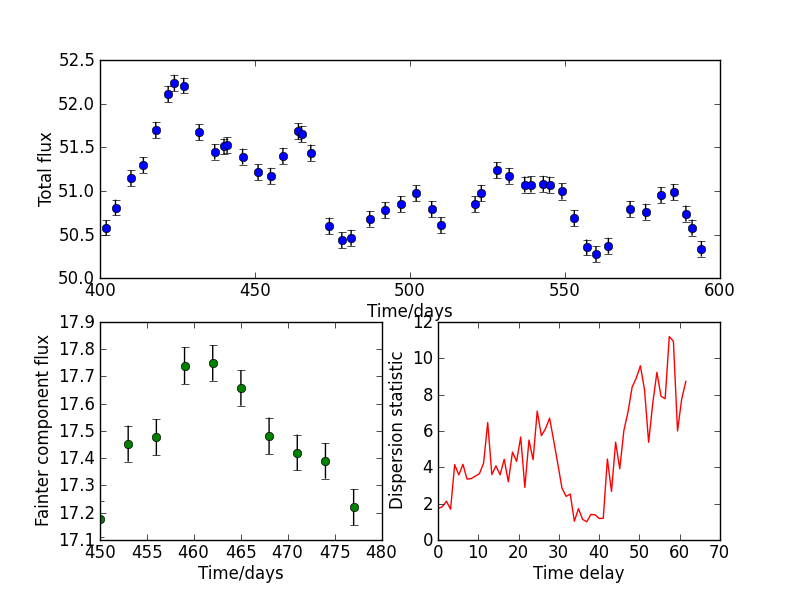}\\
\includegraphics[width=9cm]{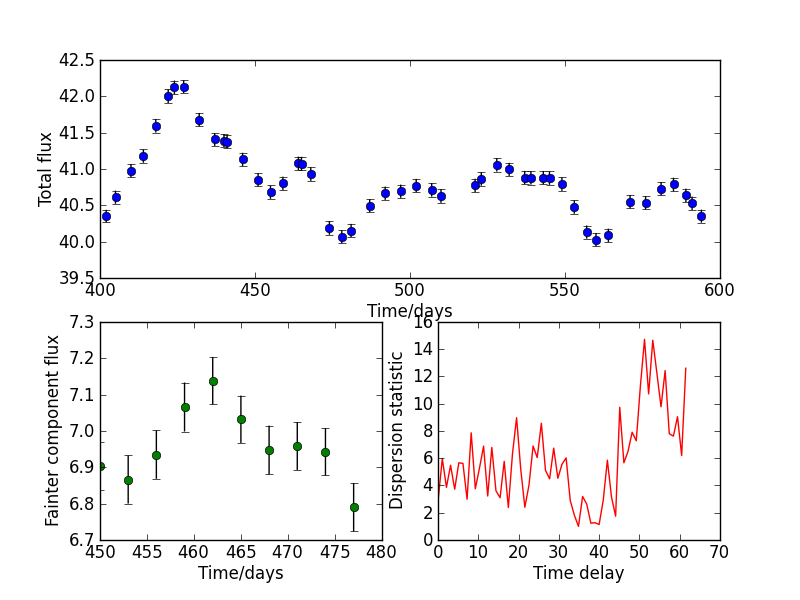}\\
\end{tabular}
\caption{Simulations of time delay recovery from two-image lenses, 
with flux ratios 2.0 (top panel) and 5.0 (bottom panel). The light curve
of CLASS~B1608+656A \citep{1999time3} has been used, and the errors
scaled so that the observation of the weaker component of the 5:1 lens
has an error of 1\%. In each panel, the top sub-panel shows the total
light curve from the two components, assuming a 36-day time delay, and
the bottom left sub-panel shows the reconstructed faint-image light curve,
adding extra Gaussian noise to the data points. The bottom right panel
shows the Pelt dispersion statistic, scaled such that $D_{\rm min}$=1.
Note that in both cases the main minimum is at approximately the correct
value, with an error of a few days in each case. There is a secondary
minimum at close to zero lag, corresponding to the influence of the fainter
component on the total light curve. In the two cases, the correct time
delays are recovered with errors of about 1 and 2.5 days for the two
cases, although this excludes a small number ($\sim$5\%) of catastrophic
errors.}\label{sec2_fig1}
\end{figure*}

A full analysis of this problem is beyond the scope of this paper, because
both the magnitude of the blending problem and the decision on triggering 
parameters depend on the particular quasar light curve as well as the
intrinsic flux ratio. However, for an illustrative example we consider the
light curve of the image A fluxes of the four-image lens system B1608+656
\citep{1999time3} as a template. This is used because it is a well
sampled light curve with a definite peak feature. We then assume a time
delay, and simulate triggered observations which compare the total-flux
light curve with a smaller number of observations of the fainter component.
The resulting total light curve, faint object light curve, and Pelt
statistic are presented in Fig. \ref{sec2_fig1} for some representative cases, and assuming a time-delay of 36 days.

\begin{figure*}
\begin{tabular}{c}
\includegraphics[width=9cm]{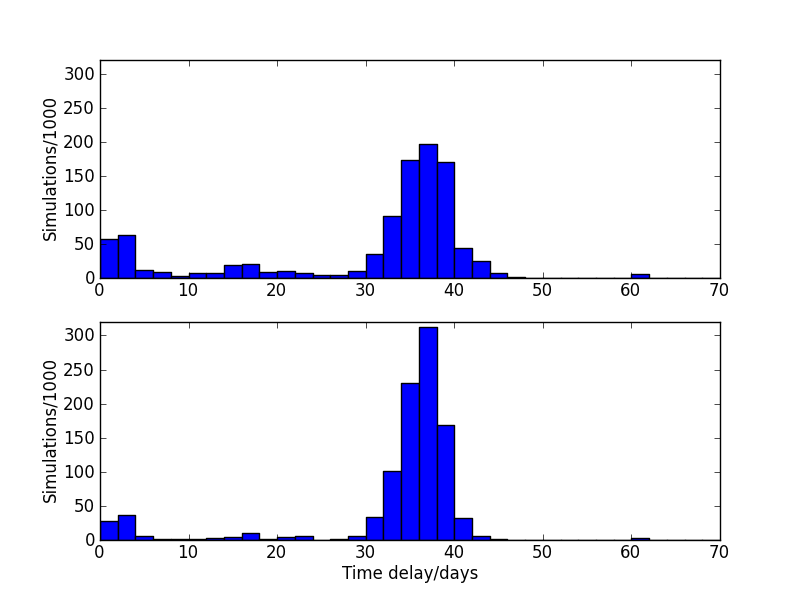}\\
\end{tabular}
\caption{Recovered time delays for the 1608+656 lightcurve, with a true 36-day time delay, but assuming a 2:1 (top) and 5:1 (bottom) flux ratio between the two components, together with a 1\% measurement error on the triggered observations of the fainter component (see fig. 1). 1000 trials were conducted in each case. Note the increased number of catastrophic errors in the 2:1 case.}\label{sec2_fig2}
\end{figure*}

As expected, the major effects on the ability to recover a good time delay
are the characteristic amplitude of variation of the quasar, as a multiple 
of the error on the observations; and the cadence of monitoring. Experiments
with different sampling of the delayed peak at 420-430 days in the total
light curve show that 10 observations, 3 days apart give an r.m.s. error
in the time delay of about half that of 3 observations, 10 days apart. For
this light curve, at least 5-8 observations are needed in order to measure
the time delay, and diminishing returns set in after this. The effect of
blending of the light curves in the total flux monitoring can also be seen
in Fig. 1. For this light curve, the effect of the fainter component on
the total light curve begins to cause a secondary minimum in the Pelt spectrum,
and significant numbers of catastrophic errors in the resultant time delay,
once the component flux ratio is about 3 or less. Fig. \ref{sec2_fig2} shows the effect on the recovered time delay as the flux ratio is lowered; for a 2:1 flux ratio there is a significant increase in the number of catastrophic errors. Again, these results are indicative only and will be different for any particular light curve.

For the preliminary observations presented in later sections, the triggering 
strategy adopted was simply to trigger further observations as soon as
a variation was seen. However, in any further observations, simulations
of this sort can and should be used to decide whether the features
seen in the total light curve give a good case for collection of
triggered observations. A further constraint which can be incorporated in
the analysis is the intrinsic ratio of the two components, if this has
previously been determined during monitoring campaigns during which the
object has not varied.

\section{WSRT 5-GHz MONITORING AND DATA REDUCTION}

An initial test of this method was made using 8 radio lenses from the CLASS
survey of gravitational lenses \citep{2003new21,2003ref02}.

\subsection{Observations and data reduction}

Total flux monitoring of 8 radio lens systems (including the highest flux-ratio double systems available in the CLASS survey) was conducted using the Westerbork
Synthesis Radio Telescope (WSRT), using observations with enough nominal sensitivity to measure variations in 
flux density at the $\sim$1\% level. Table \ref{lens} shows the observed lens systems and flux 
calibrators used in the observations. 5-GHz snapshot observations with 
8$\times$20-MHz bandwidth channels and 3\farcs7 resolution were collected 
over 39 epochs, with a separation of 2 days between epochs, from 16 July 
to 30 October 2007. Each source was observed for 10 minutes in each 
snapshot. 

The data were reduced using the National Radio Astronomy Observatory (NRAO)
Astronomical Image Processing Software ({\sc aips}) package. 3C147, a
steep-spectrum source \citep{1991new32}, was used as an absolute flux calibrator for all epochs; when
this source was not observable, the steep-spectrum source 3C138 was used instead. 
During epochs 11, 21, 31, 36 and 37 neither calibrator was observed. Therefore
observations of these epochs were excluded from the
analysis. It is worth noting that bootstrapping the flux from other sources in the field
  could not be performed because it does not yield robust results when using WSRT snapshots. The reason is that the WSRT is a linear array with a poor snapshot PSF. Each source in
each epoch was inspected separately by eye to flag bad points within 
{\sc aips} using the tasks {\sc listr, uvfnd} and {\sc uvplt} on each 
IF and Stokes parameter separately. The data of epochs 8, 27 and 34 were 
of poor quality and much of the data had to be removed, so these epochs 
were also not used for the light curves. Calibration was performed in
the standard way by a Parseltongue script which runs {\sc aips} tasks to
determine amplitude and phase solutions for the data. 

\subsection{Radio fluxes}
\begin{figure*}
\centering
\includegraphics[width=12cm,height=12cm,angle=0,keepaspectratio]{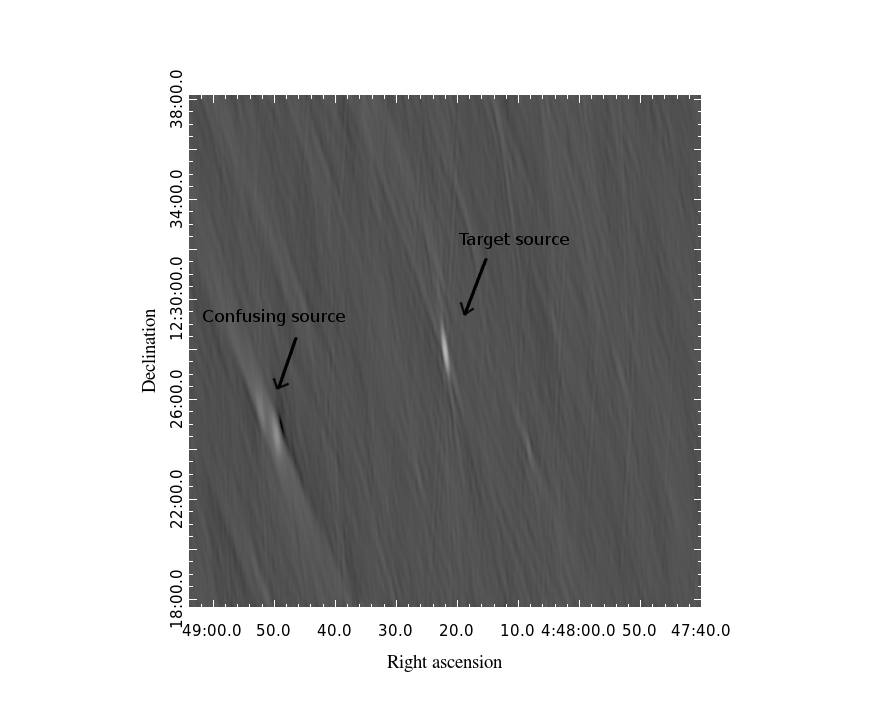}
\caption{The image presents an example for the case of a confusing source included by the beam. In the image of B0445+123, a confusing source can be seen to the southeast of the lens system. The image has been made by adding multiple epochs together to get enough UV coverage to be able to make the image. \label{confusion}}
\end{figure*}

After the calibration 
process, integrated flux densities of the target sources and the calibrators 
were derived by fitting a point-source model to the calibrated (u,v) data 
within the Caltech Difference Mapping (DIFMAP) software package \citep{1997ref9}. 
The major problem with the analysis is that the WSRT, being a linear
array, produces a fan beam which is thin and highly elongated. External
sources may be included by the beam, depending on the hour angle of 
the observation and the orientation of the target and contaminating 
sources, which may lead to a change in measured flux of a target source. Figure \ref{confusion} shows an example image of B0445+123 where a neighbouring source is included in the beam. In order to assess this effect NRAO VLA Sky Survey 
(NVSS) maps \citep{1998ref65} were obtained for each source. For 
each observation, we checked whether there are neighbouring sources 
that are likely to be included in the flux measurement by the 
orientation of the beam. We included possible confusing sources in the fitting models to subtract their effects from the fluxes measured. Further analysis was carried out to examine other possible effects such as varying the flux of contaminating sources in the models, changing the range of baselines included, and modelling the lenses as two points instead of one. These investigations showed that the NVSS sources can affect the measured fluxes. Thus, it will affect the measured scatter in the light curves. Table \ref{scatter} shows the results for the scatter measured in the light curves of the sources. In this process the NVSS sources were considered in the fitting models. Excluding short baselines in the fitting models also slightly improved our results. The chosen baselines are shown in Table\ref{scatter}. Fluxes of the NVSS sources were fixed and the lens was modelled as one point source, both of which provided better results.

The Difmap software generally produces small estimates of the error, equivalent
to the thermal noise in the maps which is expected to correspond to a measurement error on each point-source flux of $\sigma\sim0.1$ mJy. The actual scatter in the light curves is larger than this. When we examined
the light curve of 3C138, calibrated using 3C147, we found an r.m.s. scatter
of approximately 1\% (in quadrature), and we therefore adopt this as an additional error
corresponding to the best case of systematic errors in the calibration. The formal errors given by the Difmap are underestimated, since the errors are likely to be dominated by external sources intruding into the beam. These, at approximately 2\%, dominate the overall errors, and we therefore take 2\% as the error on the source flux of each measurement. It is worth noting that this is likely an overestimate for the objects that do not have a significant effect of confusing sources (e.g. B2319+051).

\subsection{Light curve production}

\begin{figure*}
\begin{center}
\scalebox{0.8}{
\begin{tabular}{cc}
\includegraphics[width=9cm,height=9cm,angle=0,keepaspectratio]{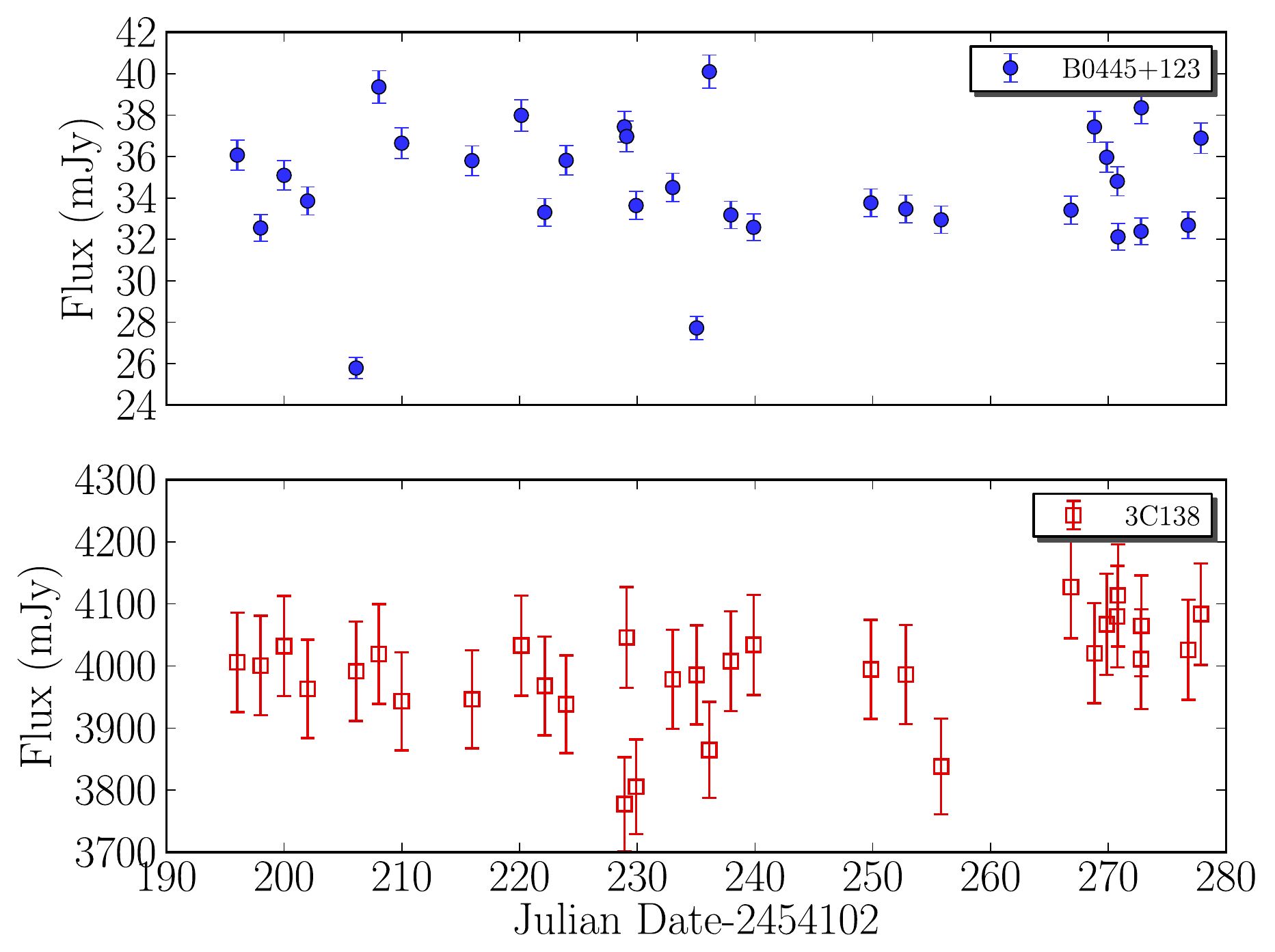}&
\includegraphics[width=9cm,height=9cm,angle=0,keepaspectratio]{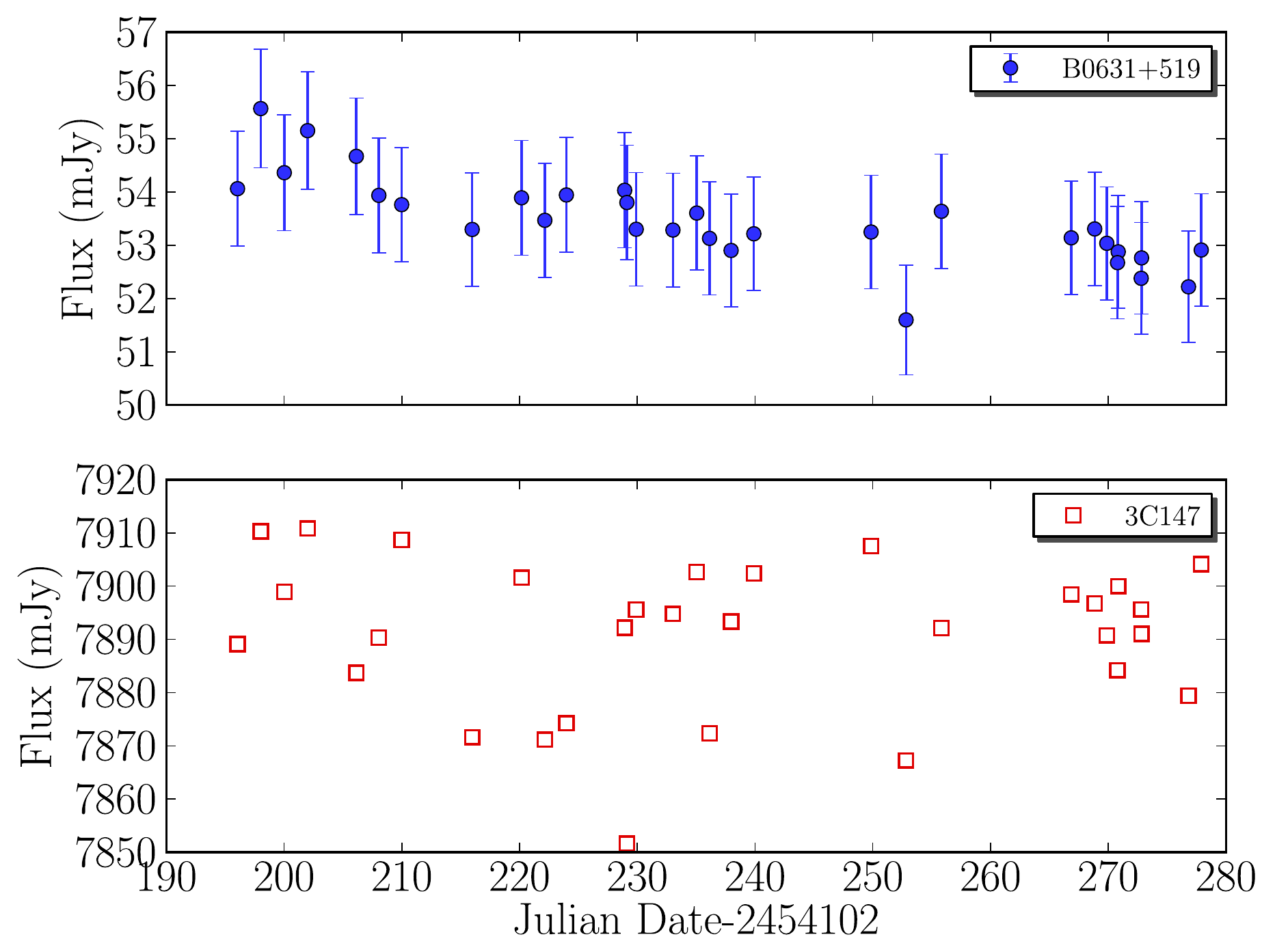}\\
\includegraphics[width=9cm,height=9cm,angle=0,keepaspectratio]{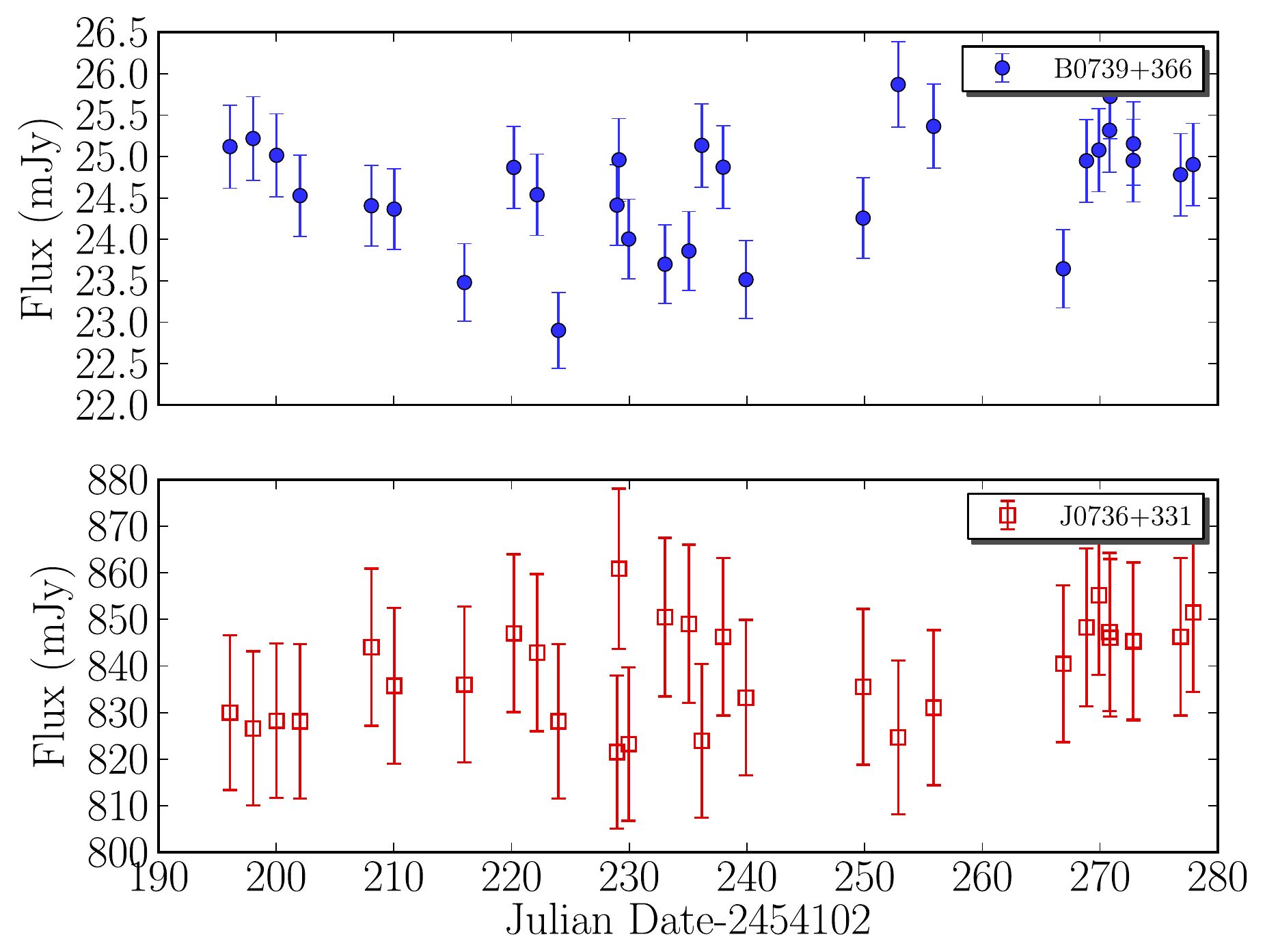}&
\includegraphics[width=9cm,height=9cm,angle=0,keepaspectratio]{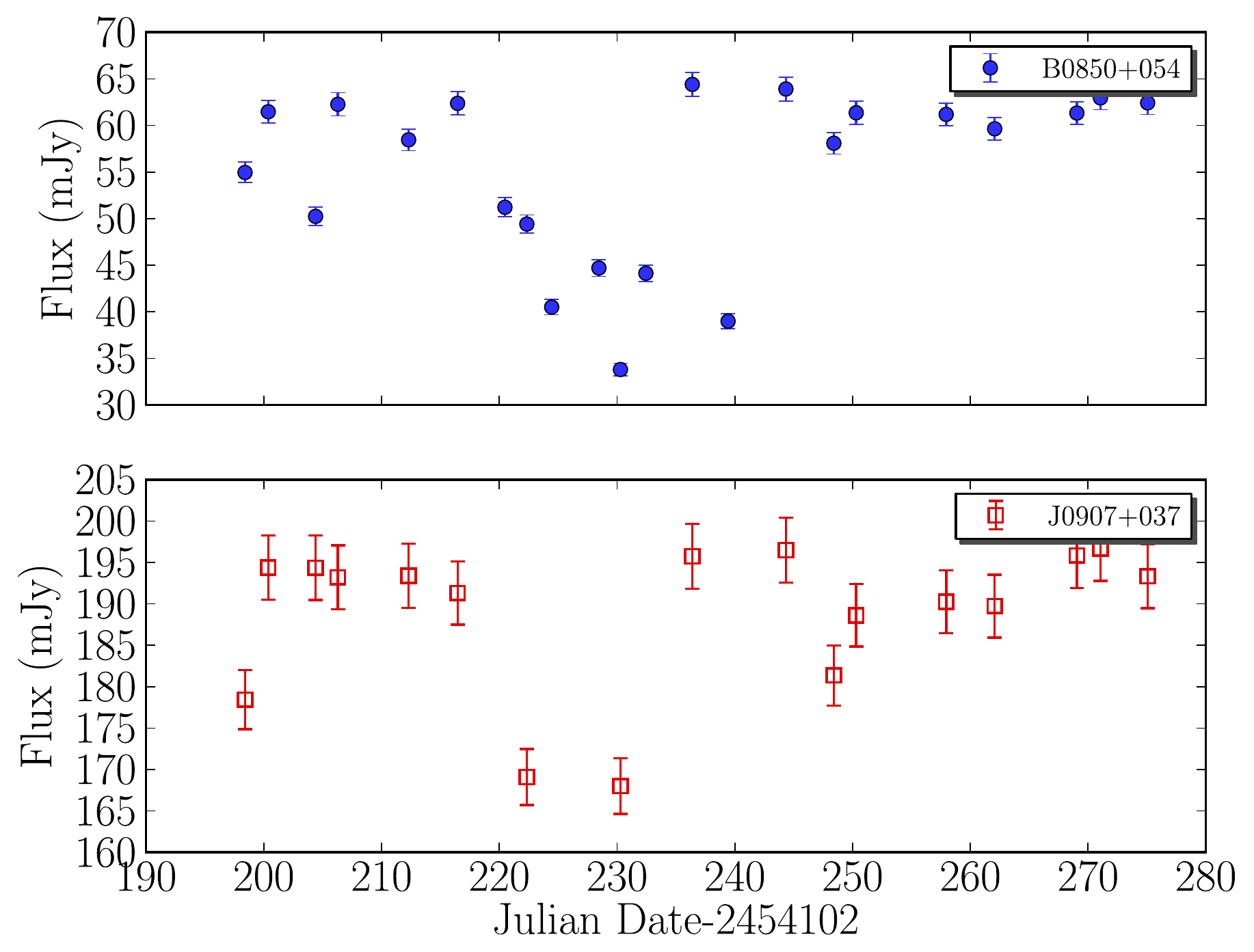}\\
\includegraphics[width=9cm,height=9cm,angle=0,keepaspectratio]{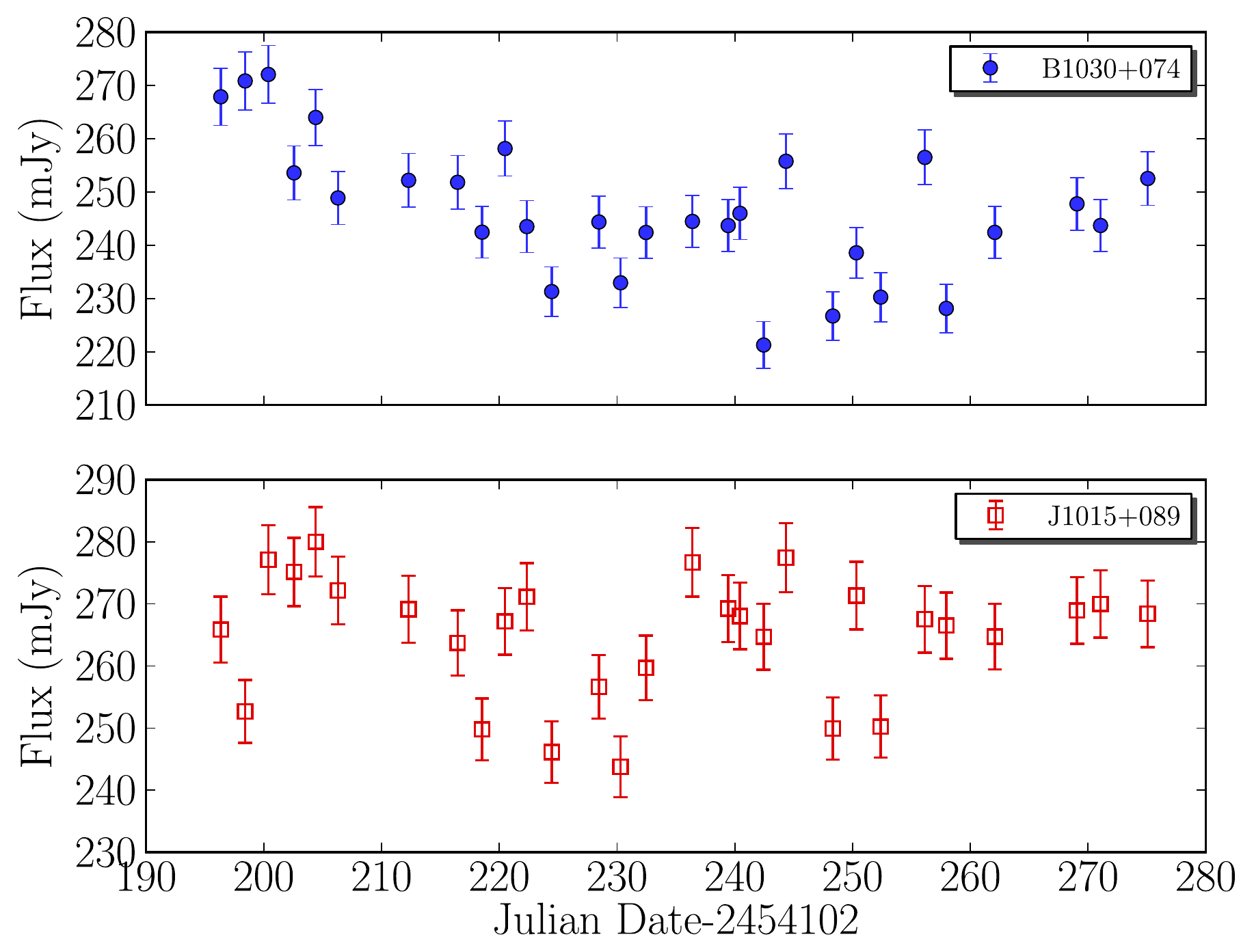}&
\includegraphics[width=9cm,height=9cm,angle=0,keepaspectratio]{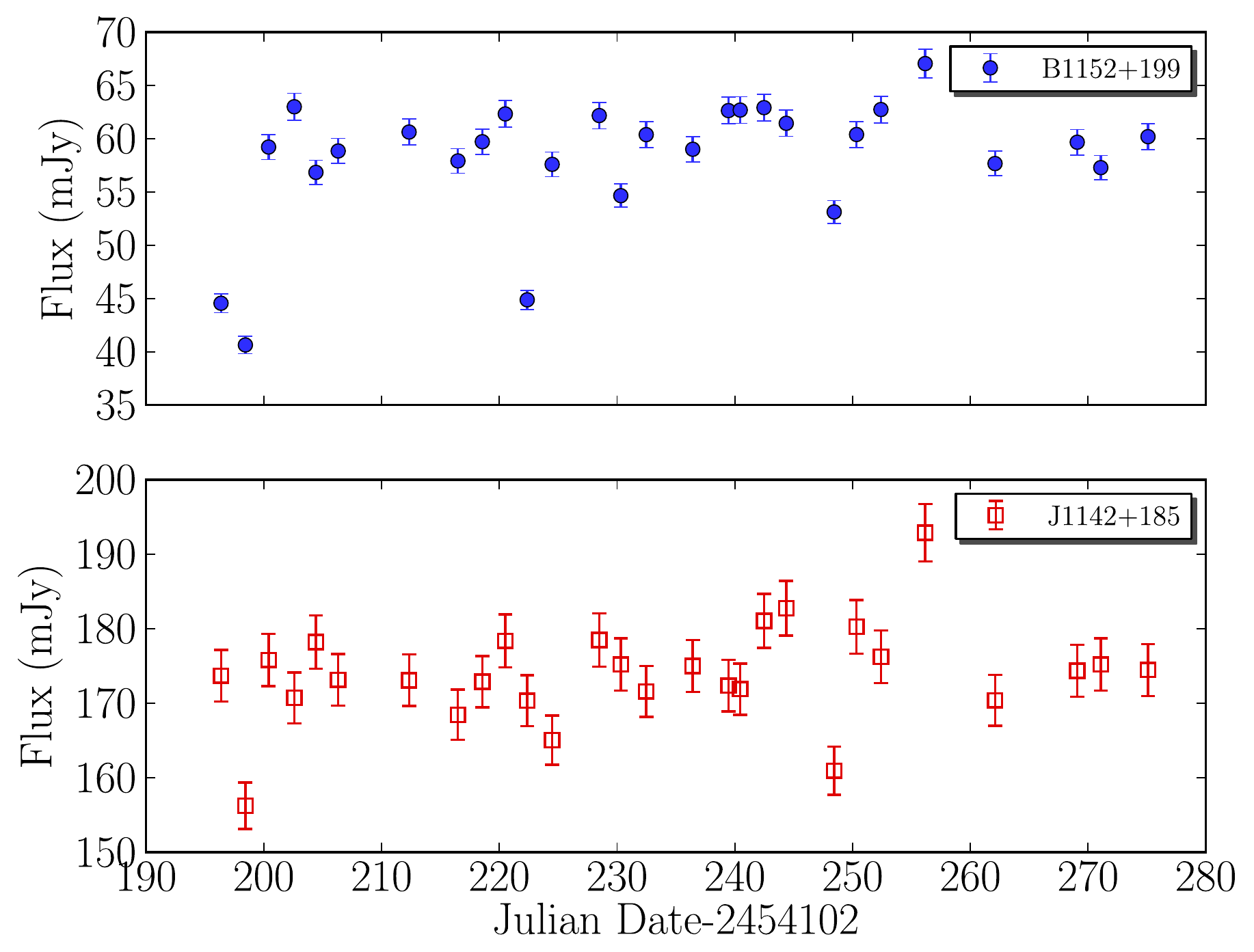}\\
\includegraphics[width=9cm,height=9cm,angle=0,keepaspectratio]{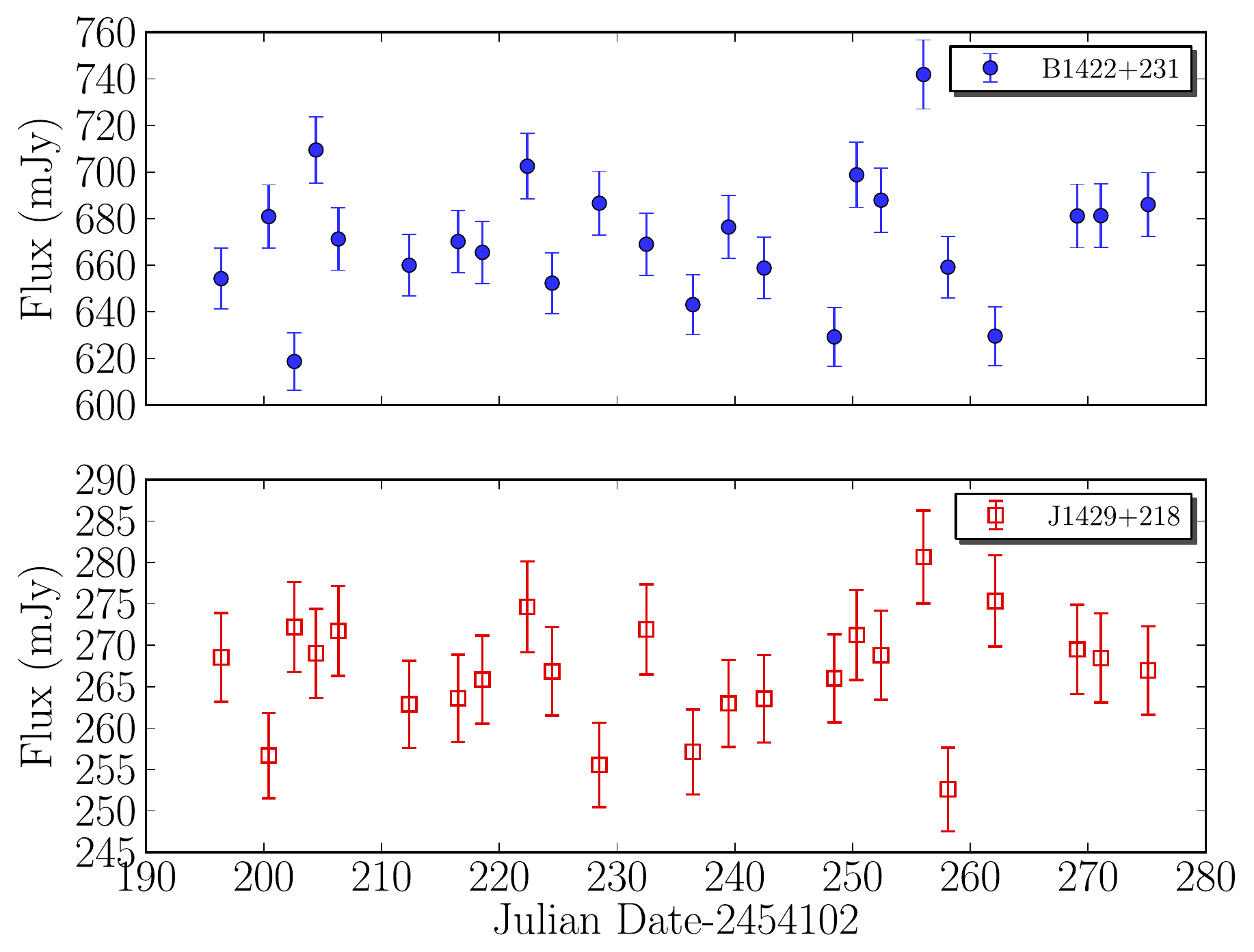}&
\includegraphics[width=9cm,height=9cm,angle=0,keepaspectratio]{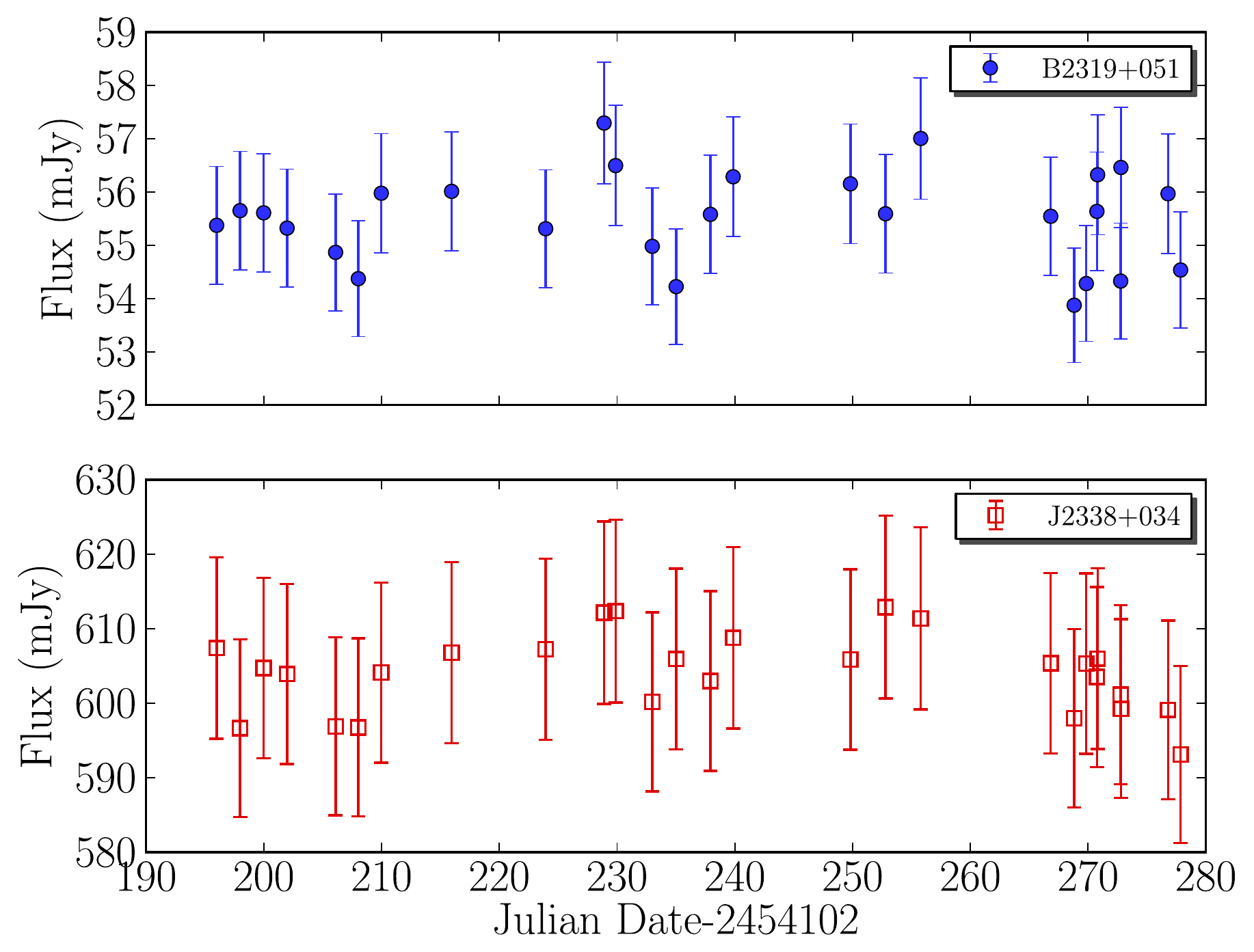}\\
\end{tabular}}
\end{center}
\caption[]{The WSRT
  light curves of the target sources (blue filled circles) with their calibrators (red open squares) plotted separately. Fluxes in the light curves were produced using the UV ranges shown in Table \ref {scatter} and NVSS sources removed from the field. Most of the calibrators' light curves are close 
  to constant (This is also checked by fitting a constant line through the calibrator fluxes and calculating the reduced $\chi^2$.) The target sources B0631+519 and B1030+074 show a variability 
  feature during the WSRT  monitoring. Some objects, in particular B0850+054, 
  had a high level of corrupted data almost all over the epochs and show a
  correspondingly greater level of scatter in their light curves. 3C147 is used as a flux calibrator and so the light curve of this source does not have errors. \label{light curves}}
\end{figure*}

Light curves of both the sources and their calibrators are shown
in Fig. \ref{light curves}. The light curves of most calibrators remain 
close to constant, although some of the calibrators are not steep-spectrum 
sources and may show intrinsic variability. A few epochs showed a significant 
scatter around the mean flux density, and in many cases this could be traced 
back to a high level of flagged data. These epochs were also removed from
subsequent analysis. Examination of Fig. \ref{light curves} shows that most 
sources, as well as calibrators, did not show obvious variability. Significant 
scatter is apparent in some sources, up to 5\% in some cases, much greater 
than the estimated errors and unlikely to be due to intrinsic variability
because of the short timescales. Among those sources whose light curve is 
reasonably smooth, B1030+074 \citep{2000new43} and B0631+519 showed possible variability 
features during the total flux monitoring. In the case of B0631+519, although the errors are uncertain, improvements in $\chi^2$ of about a factor of 2 (from 0.5 to 0.2) are obtainable by use of a straight-line fit instead of a constant flux density. This suggests that the flux density of B0631+519 varied over the monitoring period.

\begin{figure*}
\centering
\includegraphics[width=8.5cm,height=8.5cm,angle=0,keepaspectratio]{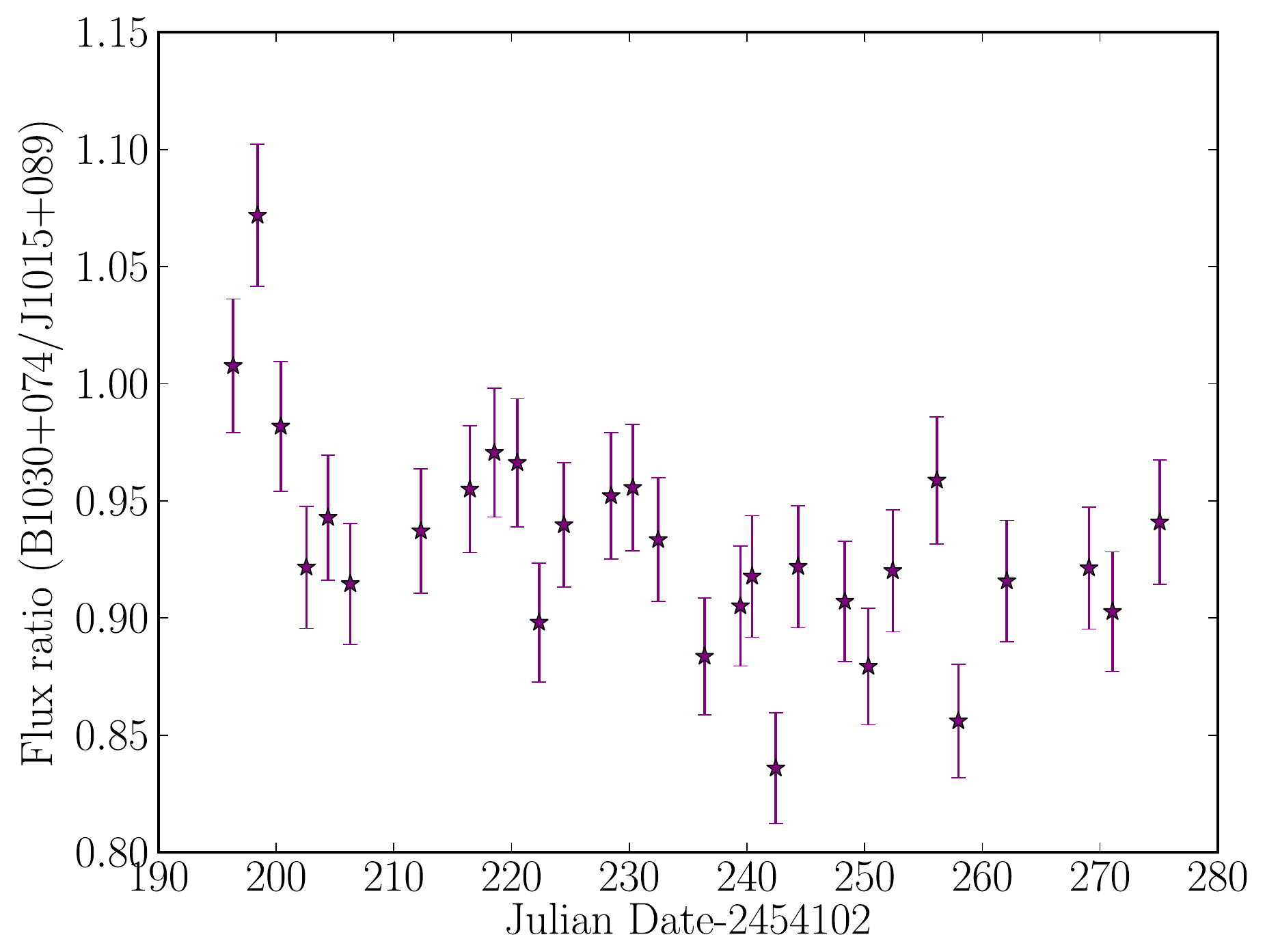}
\caption{Integrated flux densities of B1030+074,
divided by its calibrator J1015+089. Note the decrease in the flux of the target during the period of flux monitoring.\label{divided}}
\end{figure*}

In the case of B1030+074, the situation is less clear as a variation is
also apparent in the calibrator. Investigation using a Pearson statistic yields no evidence for significant correlated variation between 1030+074 and its calibrator. We conservatively assume that part of the
variation in B1030+074 is due to instrumental effects which show up in
both sources, and so Fig. \ref{divided} shows the light curve of
B1030+074 with the calibrator variation divided out. This is a well-known technique to re-normalise the light curves \citep[e.g.][]{2000time29}. A least-squares
fit to this divided curve still shows a significant reduction in
$\chi^2$ for a linear fit with a gradient, compared to that with a
constant flux; the reduced $\chi^2$ value decreases from 2.83 to 2.14.

\begin{table*}
\begin{tabular}{ccccc}
\hline
\hline
Source&Scatter&Relative&UV range&Excluded\\
&&scatter (\%)&kilo-wavelength&epochs\\
\hline
\hline
B0445+123&2.62&7.5&    8-1000&-\\
B0631+519&0.91&1.5&   13.9-1000&-\\
B0739+366&0.80&4.0&    0.0-1000&5\\
B0850+054&8.69&17.4&    0.0-1000&13,18,20,23,24,25,30,32\\
B1030+074&12.08&5.0&   0.9-4&17\\
B1152+199&5.88&11.7&   0.1-2&-\\
B1422+231&27.39&4.1&   0-1.5&3,14,19,24\\
B2319+051&1.23&2.5&    1-20&12,13,17,20\\
3C138&80.22&2.0&       0.9-1000&-\\
3C147&13.15&0.2&       10-1000&-\\
J0736+336&9.928&1.2&   0.8-10&-\\
J0907+037&8.41&4.6&   10-20&5,13,17,24,25,30,32\\
J1015+089&9.53&3.7&    0.8-4&17\\
J1142+185&6.73&3.6&    1.5-10&-\\
J1429+218&6.32&2.4&    10-13&7,33,35\\
J2338+034&6.41&1.1&    10-20&12,13,17,20\\
\hline
\end{tabular}
\caption[Scatters]{The table shows the r.m.s. scatters measured in the light curves of all sources. The units are in mJy.} 
\label{scatter}
\end{table*}

\section{VLA 5-GHz MONITORING AND DATA REDUCTION}

During the total flux monitoring only B1030+074 and
B0631+519 target sources showed a possible variability feature
in their light curves. As soon as the variability feature was seen in the first 13 epochs ($\sim$ 220 day), VLA observations were triggered only for B1030+074 at nine epochs
(separated by 10 days) between 
2007 October 5 and 2008 January 3. The VLA observing period was chosen to
be around the time that we expect the fainter component to vary, provided 
that $H_{0}$ is between 50 and 100 km s$^{-1}$ Mpc$^{-1}$ and that the lens
galaxy's mass profile is not too far from isothermal \citep{2006method17,2009referans49,2011ref50}. Because B0631+519
is a smaller-separation system, the variation which occurred late in the observing session could not be followed up with the VLA.

For each epoch, the data were collected at 5 GHz using 2 IFs each
with 50 MHz bandwidth. The array completed a move from the extended A-configuration
to B-configuration during this time, so the observations were conducted
using B-configuration, which has a resolution of 1\farcs2, just sufficient
to separate, and measure the flux densities of the two components of B1030+074 
which are 1\farcs6 apart \citep{1998ref5}. 3C286 was used as a primary
amplitude calibrator and 1143+185, a compact source that does not vary
\citep{2001ref48}, was used as a secondary calibrator to improve errors in the
absolute flux calibration for each epoch. 1015+089 and 1014+088 were used as
phase calibrators in different epochs. In each epoch, the target source was
observed first, followed by the phase calibrator, secondary flux calibrator
and primary flux calibrator respectively. 

The NRAO {\sc aips} package was used for initial editing and all of the calibration
processes. Before gain calibration, the data were manually inspected, and if
necessary flagged. The first epoch had large scatter in the visibility amplitudes 
and it was not used for the subsequent analysis. 

During the observations the EVLA antennas were included. The mismatch between the bandpass response functions of VLA and
EVLA antennas can lead to closure errors \footnote{http://www.vla.nrao.edu/astro/guides/evlareturn/vla-evla.shtml/}. These errors were removed
at the beginning of the analysis using the observations of 3C286 as
a baseline calibrator. These baseline calibration results were
applied to the data during the subsequent gain calibration process for all 
epochs. Calibration proceeded by obtaining amplitude and phase solutions
for each antenna and applying these to the data. Finally, imaging of
the target source B1030+074 and self-calibration was carried out manually
for each epoch.

\subsection{Light curve production}

\begin{figure*}
\begin{center}
\scalebox{1}{
\begin{tabular}{cc}
\includegraphics[width=6cm,angle=0]{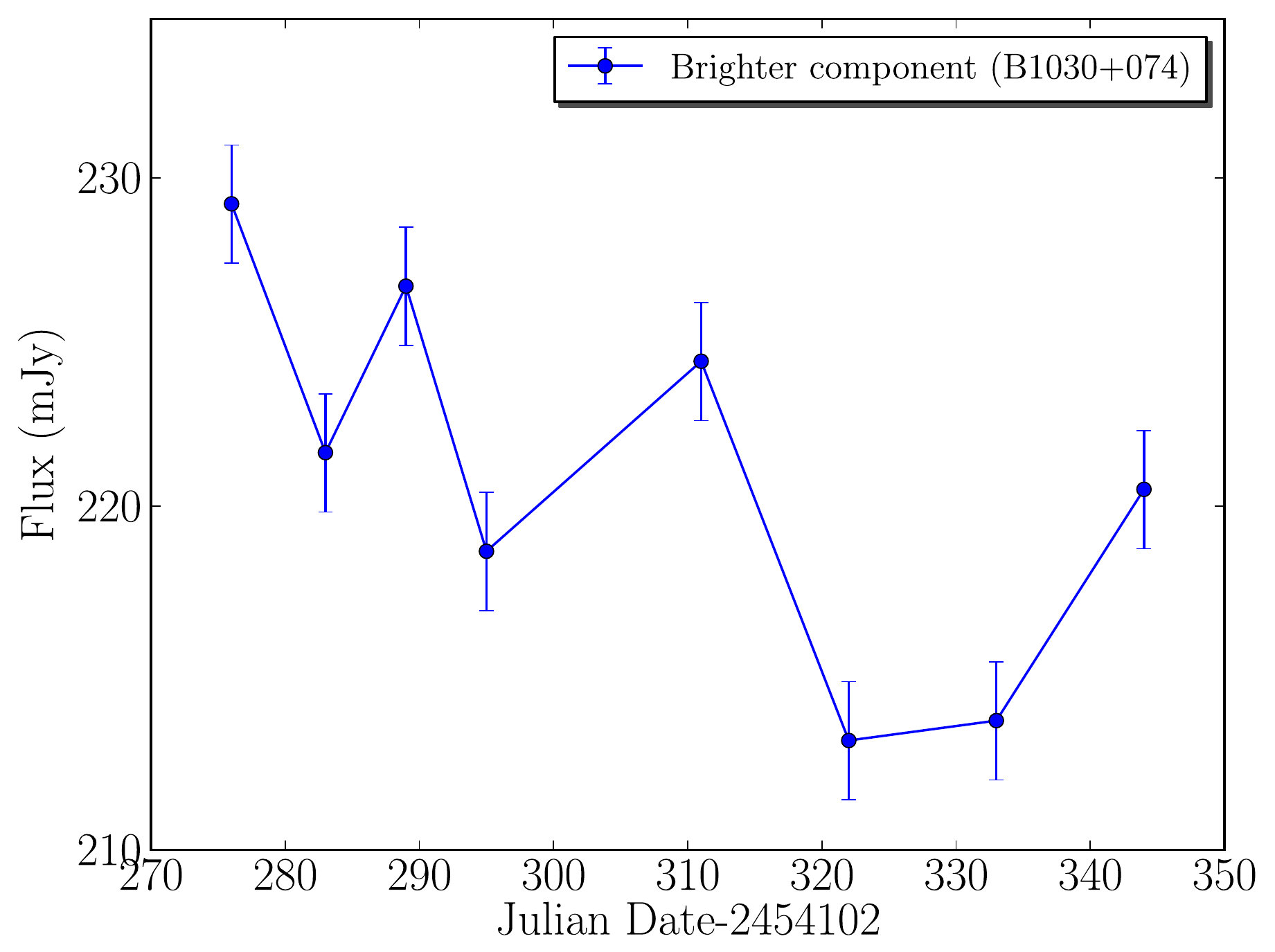}&
\includegraphics[width=6cm,angle=0]{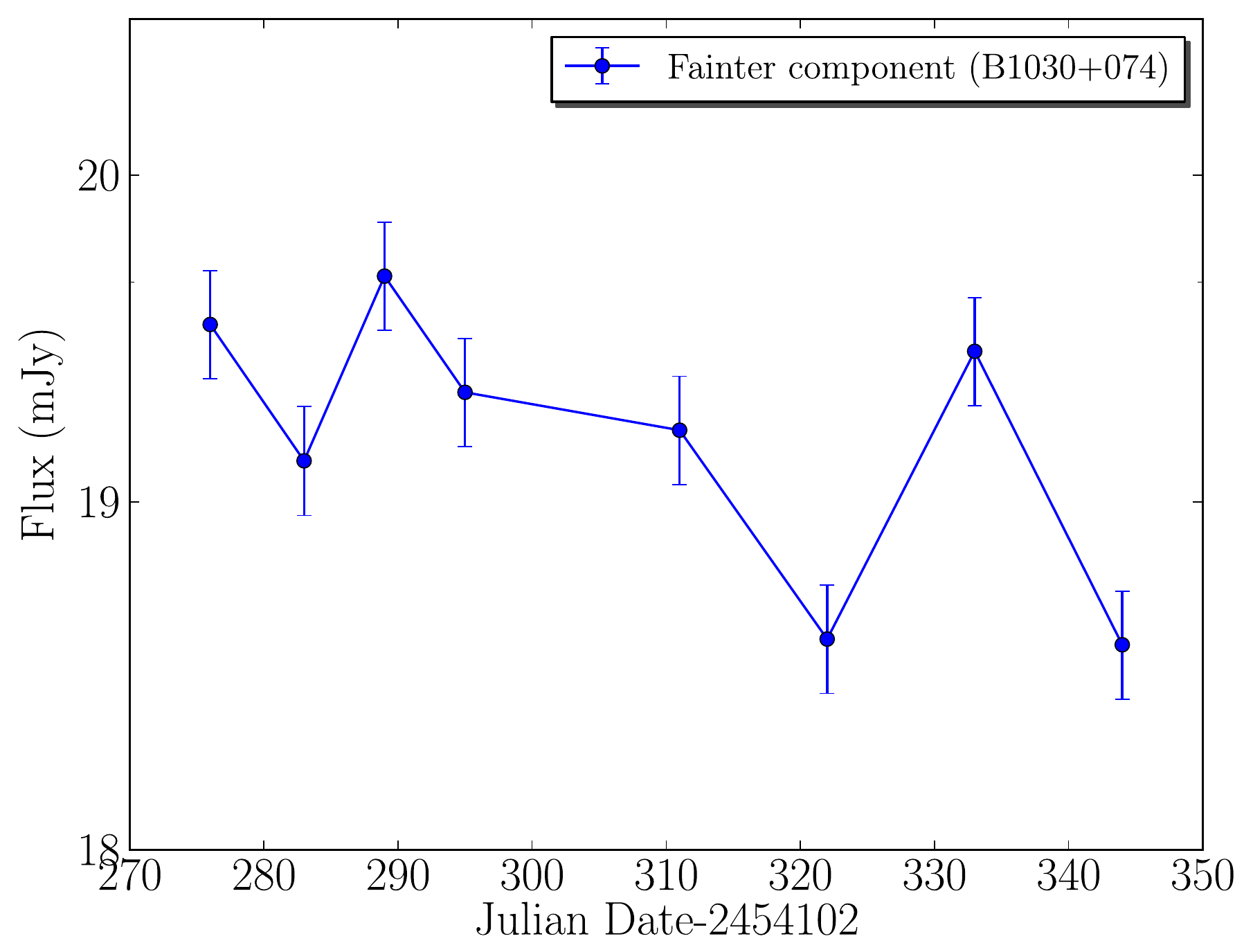}\\
\includegraphics[width=6.65cm,angle=0]{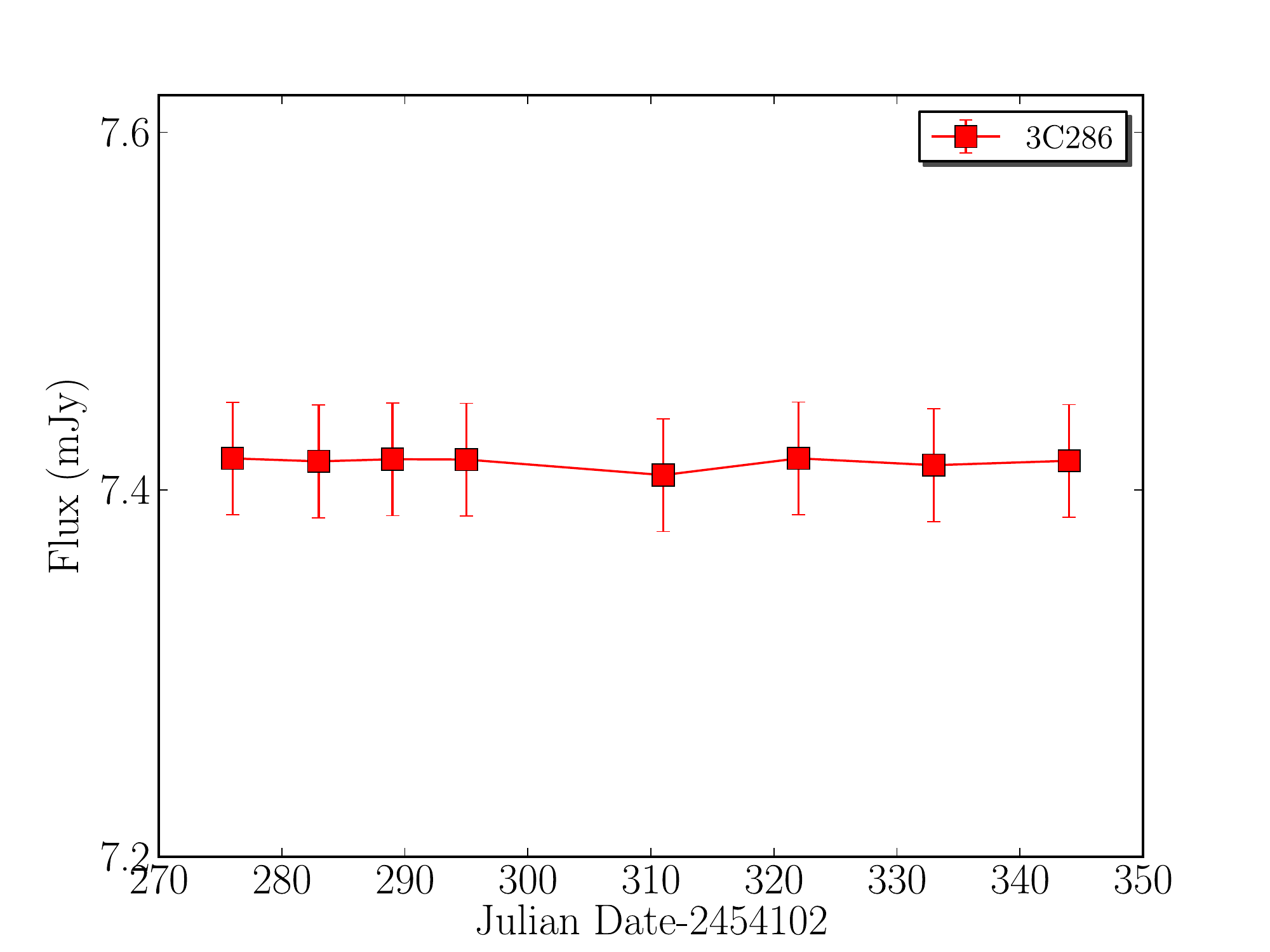}&
\includegraphics[width=6.65cm,angle=0]{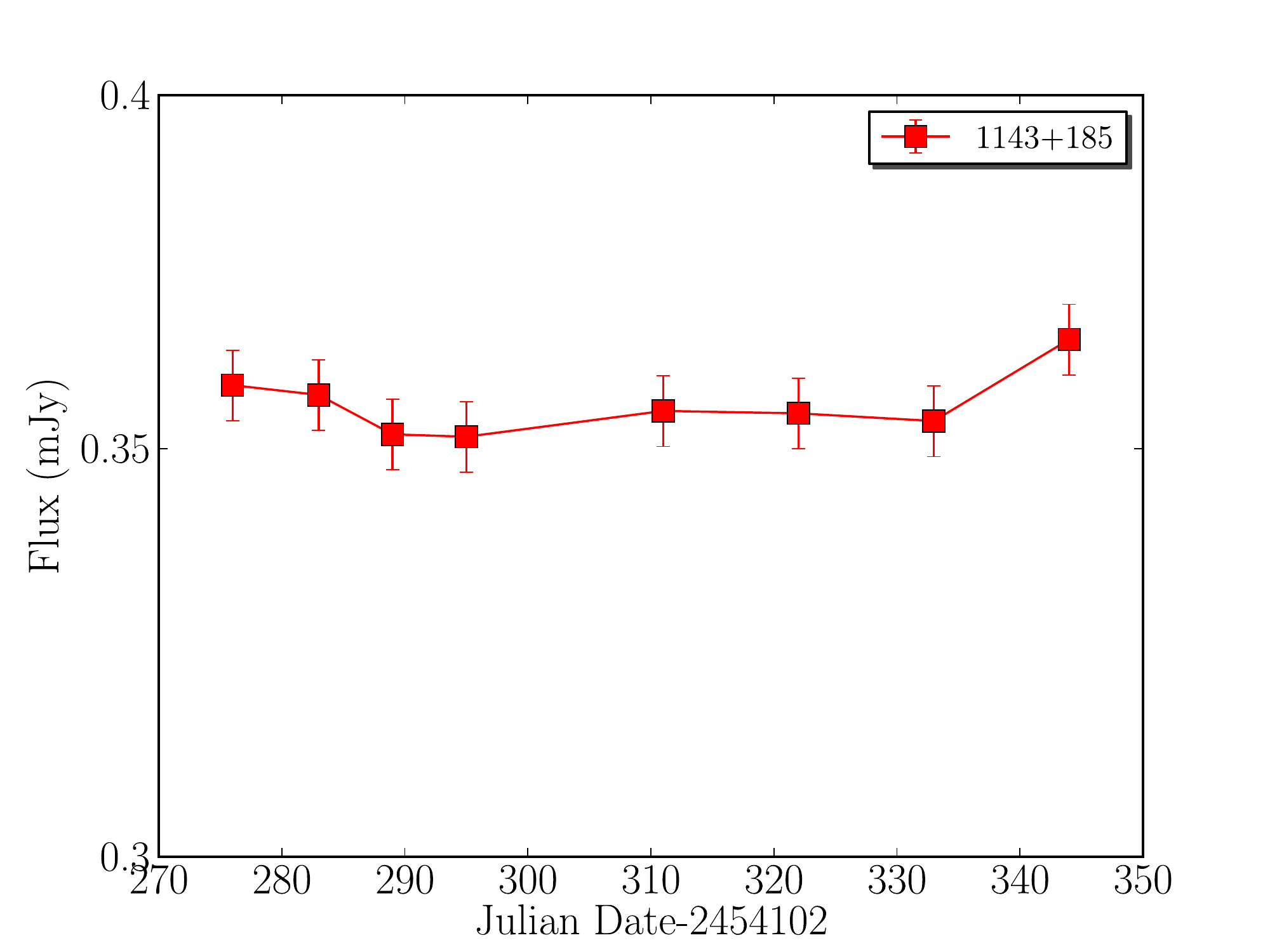}\\
\end{tabular}}
\end{center}
\caption{VLA light curves for the triggered observations of the B1030+074 and the flux calibrators. The first top plot: the brighter component. The second plot: the fainter component shows a declining trend. For comparison, the bottom two figures show the light curves of the two calibrators 3C286 (assumed to have a constant flux density) and the flux calibrator 1143+185, which is a non-variable compact source.}
\label{vla1030}
\end{figure*}

Flux densities of the components of B1030+074 were derived by fitting a two
point-source model to the ($\textit{u,v}$) data within the {\sc difmap} 
software package, with variable fluxes but with known relative positions \citep{1998ref5}. The 
flux calibrators were processed in the same manner to measure
their flux densities. Corresponding light curves were produced using
these flux density values. Error estimations of the fluxes, $\sigma_{flux}$ 
were calculated using \citep{2004ref66};

\begin{equation}
\sigma_{flux} = \sqrt{(\sigma_{\rm{thermal}})^2+(\sigma_{\rm{c}} \times \textrm
{peak flux density})^2}, 
\end{equation}

where $\sigma_{\rm{thermal}}$ is the thermal noise.
This estimation takes into account errors of statistical noise and point
source calibrations. A thermal noise (statistical noise) was determined by
measuring the r.m.s. background noise far away from the source. The calibration error, denoted as $\sigma_{\rm{c}}$, which depends on the accuracy of the adopted flux density scale and is usually of the order of a few per cent \citep{1977ref75}, was introduced into the total error
estimation by adding 2$\%$ in quadrature as a conservative approach. Otherwise, the errors derived from
either {\sc aips} or {\sc difmap} are underestimated since these programmes 
calculate only the difference between observed and model visibilities for 
all data points. The light curves of the brighter and fainter component
separately, are shown in Fig. \ref{vla1030}. There is some suggestion that
the brighter component continues to decline in flux density. Our main interest,
however, is in the variation of the faint component, as any change in this
component's brightness should reflect the decline previously seen in the
total flux density of the source (which is dominated by the bright component).
In fact, fitting to this light curve using a linear function of arbitrary 
gradient provided some improvement (reduced $\chi^2$=1.14) in comparison to fitting 
a constant flux density (reduced $\chi^2$=1.6), although the data are noisy enough that the latter cannot be ruled out.

\subsection{Analysis of the light curves and time delay estimation}

\begin{figure*}
\begin{tabular}{cc}
\includegraphics[width=8.5cm]{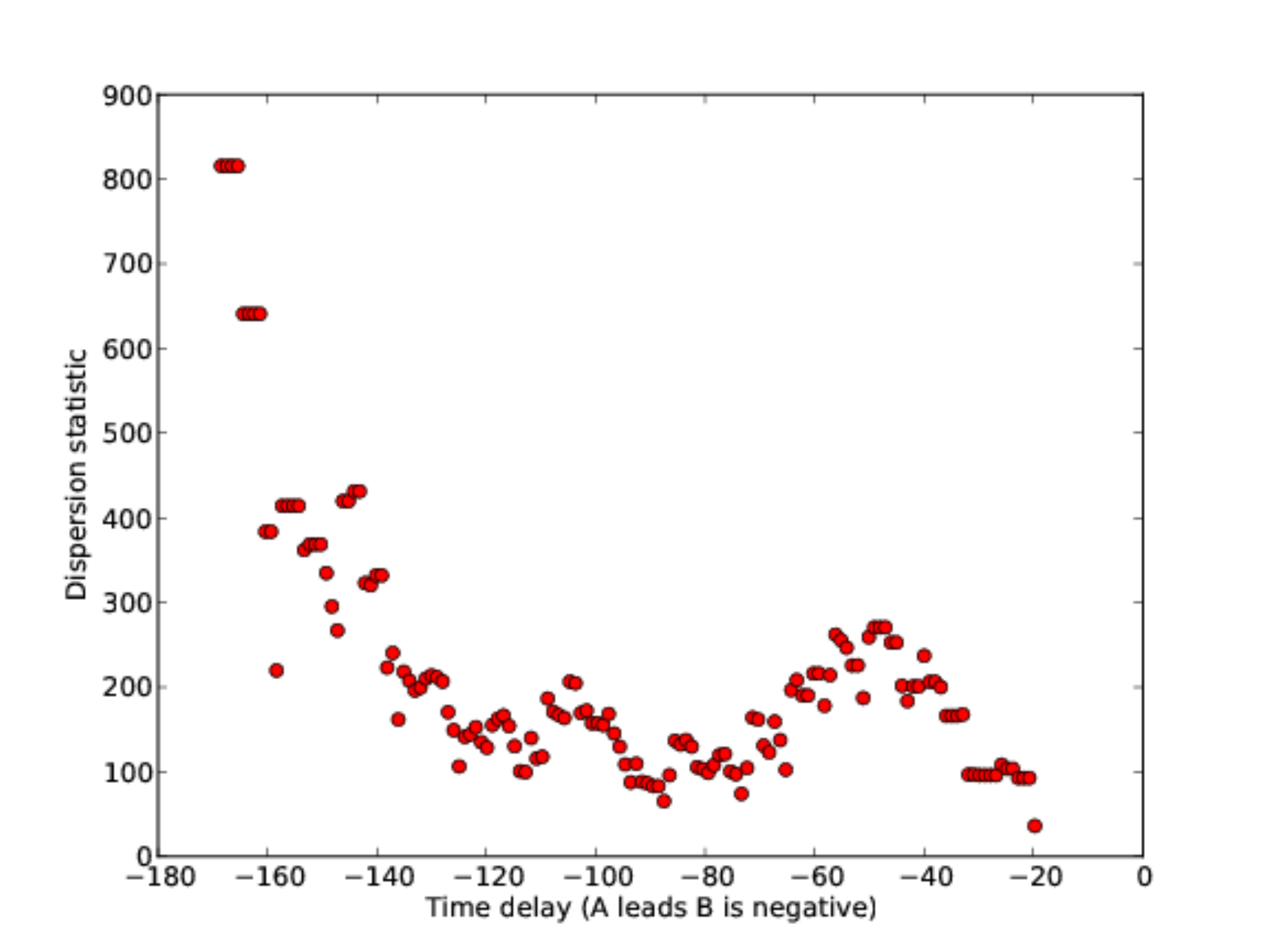}&
\includegraphics[width=8.5cm]{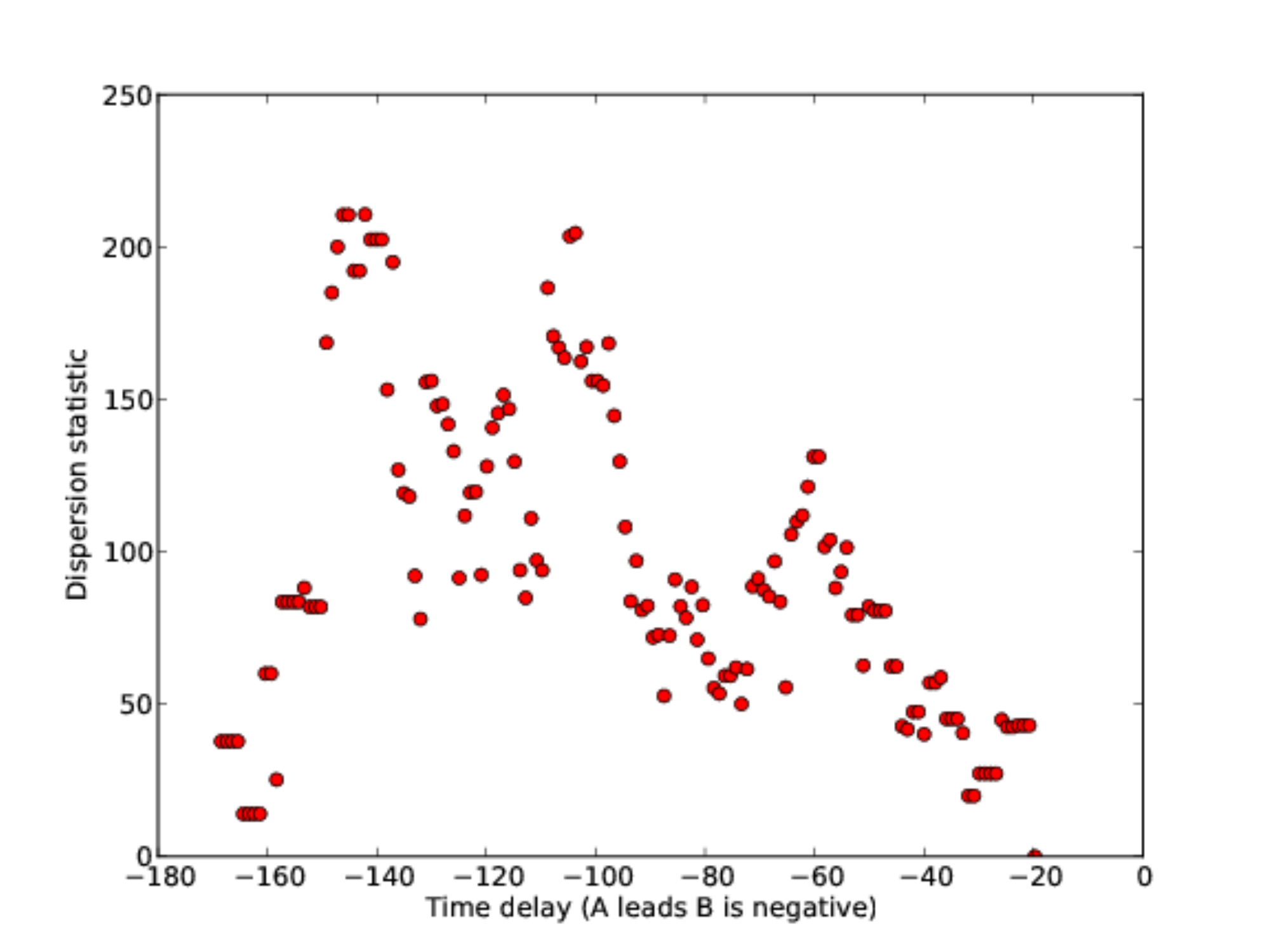}\\
\end{tabular}
\caption{Inferred Pelt dispersion statistic for the
WSRT and VLA observations of B1030+074. Left: assuming that the intrinsic
flux ratio is known to be 13 \citep[e.g.][]{1998ref5}. Right: with
no assumption about the intrinsic flux ratio. It is worth noting that the dispersion statistic at the extreme left-hand end of the plot ($<160$ d) is based on the overlap of very few points.}
\label{1030_pelt}
\end{figure*}

Using the VLA and WSRT light curves together, we calculated
values for the Pelt statistic as a function of time delay. The 
results are shown in Fig. \ref{1030_pelt}. We cannot extract
an unambiguous time delay, because of a number of circumstances including
a lack of a clear peak in the flux density variation, and the relatively
high scatter in the followup observations, which may be due to difficulties
in the amplitude calibration owing to the VLA/EVLA changeover which was
happening during the observations. The results do show a wide minimum
in the possible time delay. If we assume a known flux density ratio between the
two components, this rules out very long time delays but otherwise does
not improve the constraints significantly. In theory, knowing the intrinsic 
flux density ratio should allow a time delay determination if only a monotonic 
decline was observed in the source, but this would require higher signal-to-noise 
than is available in the observations reported here. The predicted time delay of 
$\sim$110 days \citep{1998ref5}, expected if $H_0=70\,$km$\,$s$^{-1}$$\,$Mpc$^{-1}$, is consistent with the
data.

\section{CONCLUSIONS}
In this work we have proposed an alternative method for gravitational lens time delay measurements. This technique does not rely only on high-resolution observations which are typically required for lensing time delay measurements. It primarily uses low-resolution observations and this enables us to utilise high-resolution observations at an optimum level.

The efficiency of this technique, defined as the number of high-resolution observations that it requires, was evaluated by performing cross-correlation simulations using the Pelt dispersion statistic. Our results show that, for typical lightcurves, the true time delay can be covered with 5-8 high-resolution observations, an order of magnitude fewer than required in traditional approaches. As a pilot project, we used
the WSRT to perform total flux monitoring for 8 radio lens systems and
triggered VLA observations for the one object, B1030+074, that showed 
variability during the total flux monitoring. For this object, the expected 
trend of decreasing flux density with time was not seen convincingly in the 
fainter component's light curve. Analysis of the possible time delay concluded 
that a wide range of time delays are consistent with the available data.

Despite the lack of a clear result on an initial trial, 
this new method is potentially useful because it predominantly
uses time on low-resolution telescopes. This is important because new, highly sensitive but low-resolution instruments are under construction such as MeerKAT (an RMS noise level of $\sim$ 7$\mu$Jy/beam in 24 hours with 500 MHz)
and ASKAP (an RMS noise level of $\sim$ 37$\mu$Jy/beam in an hour with 300 MHz). Since these arrays are not linear, confusion due to neighbouring sources will not be a big problem. Such instruments, together with a modest amount
of high-resolution observational followup, may in future be useful
for gravitational lensing time delay measurements by means of this
new method.

\section*{Acknowledgements}

We would like to thank Ian Browne for his useful comments and discussions. The Westerbork Synthesis Radio Telescope is operated by the ASTRON 
(Netherlands Institute for Radio Astronomy) with support from the 
Netherlands Foundation for Scientific Research (NWO). We thank Michiel
Brentjens for support during the WSRT observations. 
The Very Large Array is operated by the the U.S. National Radio Astronomy 
Observatory, which is a facility of the National Science Foundation operated 
under cooperative agreement by Associated Universities, Inc.

\bibliographystyle{mn2e}
\bibliography{all_paper}

\begin{thebibliography}{59}
\expandafter\ifx\csname natexlab\endcsname\relax\def\natexlab#1{#1}\fi

\bibitem[{{Argo} {et~al}\mbox{.}(2003){Argo}, {Jackson}, {Browne}, {York},
  {McKean}, {Biggs}, {Blandford}, {de Bruyn}, {Chae}, {Fassnacht}, {Koopmans},
  {Myers}, {Norbury}, {Pearson}, {Phillips}, {Readhead}, {Rusin}, \&
  {Wilkinson}}]{2003ref1}
{Argo} M.~K. {et~al.}, 2003, \mnras, 338, 957

\bibitem[{{Auger} {et~al}\mbox{.}(2010){Auger}, {Treu}, {Bolton}, {Gavazzi},
  {Koopmans}, {Marshall}, {Moustakas}, \& {Burles}}]{2010method20}
{Auger} M.~W., {Treu} T., {Bolton} A.~S., {Gavazzi} R., {Koopmans} L.~V.~E.,
  {Marshall} P.~J., {Moustakas} L.~A., {Burles} S., 2010, \apj, 724, 511

\bibitem[{{Baars} {et~al}\mbox{.}(1977){Baars}, {Genzel}, {Pauliny-Toth}, \&
  {Witzel}}]{1977ref75}
{Baars} J.~W.~M., {Genzel} R., {Pauliny-Toth} I.~I.~K., {Witzel} A., 1977,
  \aap, 61, 99

\bibitem[{{Barnab{\`e}} {et~al}\mbox{.}(2011){Barnab{\`e}}, {Czoske},
  {Koopmans}, {Treu}, \& {Bolton}}]{2011ref50}
{Barnab{\`e}} M., {Czoske} O., {Koopmans} L.~V.~E., {Treu} T., {Bolton} A.~S.,
  2011, \mnras, 415, 2215

\bibitem[{{Barnab{\`e}} \& {Koopmans}(2007)}]{2007new42}
{Barnab{\`e}} M., {Koopmans} L.~V.~E., 2007, \apj, 666, 726

\bibitem[{{Biggs} {et~al}\mbox{.}(2001){Biggs}, {Browne}, {Wilkinson},
  {Muxlow}, {Helbig}, \& {Koopmans}}]{2001method18}
{Biggs} A.~D., {Browne} I.~W.~A., {Wilkinson} P.~N., {Muxlow} T.~W.~B.,
  {Helbig} P., {Koopmans} L.~V.~E., 2001, in Astronomical Society of the
  Pacific Conference Series, Vol. 237, Gravitational Lensing: Recent Progress
  and Future Go, {T.~G.~Brainerd \& C.~S.~Kochanek}, ed., p. 137

\bibitem[{{Biggs} {et~al}\mbox{.}(2003){Biggs}, {Rusin}, {Browne}, {de Bruyn},
  {Jackson}, {Koopmans}, {McKean}, {Myers}, {Blandford}, {Chae}, {Fassnacht},
  {Norbury}, {Pearson}, {Phillips}, {Readhead}, \& {Wilkinson}}]{2003ref4}
{Biggs} A.~D. {et~al.}, 2003, \mnras, 338, 1084

\bibitem[{{Browne} {et~al}\mbox{.}(2003){Browne}, {Wilkinson}, {Jackson},
  {Myers}, {Fassnacht}, {Koopmans}, {Marlow}, {Norbury}, {Rusin}, {Sykes},
  {Biggs}, {Blandford}, {de Bruyn}, {Chae}, {Helbig}, {King}, {McKean},
  {Pearson}, {Phillips}, {Readhead}, {Xanthopoulos}, \& {York}}]{2003ref02}
{Browne} I.~W.~A. {et~al.}, 2003, \mnras, 341, 13

\bibitem[{{Chang} \& {Refsdal}(1979)}]{1979ref90}
{Chang} K., {Refsdal} S., 1979, \nat, 282, 561

\bibitem[{{Condon} {et~al}\mbox{.}(1998){Condon}, {Cotton}, {Greisen}, {Yin},
  {Perley}, {Taylor}, \& {Broderick}}]{1998ref65}
{Condon} J.~J., {Cotton} W.~D., {Greisen} E.~W., {Yin} Q.~F., {Perley} R.~A.,
  {Taylor} G.~B., {Broderick} J.~J., 1998, \aj, 115, 1693

\bibitem[{{Courbin} {et~al}\mbox{.}(2011){Courbin}, {Chantry}, {Revaz},
  {Sluse}, {Faure}, {Tewes}, {Eulaers}, {Koleva}, {Asfandiyarov}, {Dye},
  {Magain}, {van Winckel}, {Coles}, {Saha}, {Ibrahimov}, \&
  {Meylan}}]{2011new25}
{Courbin} F. {et~al.}, 2011, \aap, 536, A53

\bibitem[{{Dobke} {et~al}\mbox{.}(2009){Dobke}, {King}, {Fassnacht}, \&
  {Auger}}]{2009new29}
{Dobke} B.~M., {King} L.~J., {Fassnacht} C.~D., {Auger} M.~W., 2009, \mnras,
  397, 311

\bibitem[{{Eigenbrod} {et~al}\mbox{.}(2005){Eigenbrod}, {Courbin}, {Vuissoz},
  {Meylan}, {Saha}, \& {Dye}}]{2005new22}
{Eigenbrod} A., {Courbin} F., {Vuissoz} C., {Meylan} G., {Saha} P., {Dye} S.,
  2005, \aap, 436, 25

\bibitem[{{El{\'{\i}}asd{\'o}ttir}
  {et~al}\mbox{.}(2006){El{\'{\i}}asd{\'o}ttir}, {Hjorth}, {Toft}, {Burud}, \&
  {Paraficz}}]{2006ref88}
{El{\'{\i}}asd{\'o}ttir} {\'A}., {Hjorth} J., {Toft} S., {Burud} I., {Paraficz}
  D., 2006, \apjs, 166, 443

\bibitem[{{Fassnacht} {et~al}\mbox{.}(1999){Fassnacht}, {Pearson}, {Readhead},
  {Browne}, {Koopmans}, {Myers}, \& {Wilkinson}}]{1999time3}
{Fassnacht} C.~D., {Pearson} T.~J., {Readhead} A.~C.~S., {Browne} I.~W.~A.,
  {Koopmans} L.~V.~E., {Myers} S.~T., {Wilkinson} P.~N., 1999, \apj, 527, 498

\bibitem[{{Fassnacht} \& {Taylor}(2001)}]{2001ref48}
{Fassnacht} C.~D., {Taylor} G.~B., 2001, \aj, 122, 1661

\bibitem[{{Fassnacht} {et~al}\mbox{.}(2002){Fassnacht}, {Xanthopoulos},
  {Koopmans}, \& {Rusin}}]{2002time4}
{Fassnacht} C.~D., {Xanthopoulos} E., {Koopmans} L.~V.~E., {Rusin} D., 2002,
  \apj, 581, 823

\bibitem[{{Fohlmeister} {et~al}\mbox{.}(2008){Fohlmeister}, {Kochanek},
  {Falco}, {Morgan}, \& {Wambsganss}}]{2008new27}
{Fohlmeister} J., {Kochanek} C.~S., {Falco} E.~E., {Morgan} C.~W., {Wambsganss}
  J., 2008, \apj, 676, 761

\bibitem[{{Freedman} \& {Madore}(2010)}]{2010ref64}
{Freedman} W.~L., {Madore} B.~F., 2010, \araa, 48, 673

\bibitem[{{Geiger} \& {Schneider}(1996)}]{1996method1}
{Geiger} B., {Schneider} P., 1996, \mnras, 282, 530

\bibitem[{{Homan} {et~al}\mbox{.}(2004){Homan}, {Wijnands}, {Rupen}, {Fender},
  {Hjellming}, {di Salvo}, \& {van der Klis}}]{2004ref66}
{Homan} J., {Wijnands} R., {Rupen} M.~P., {Fender} R., {Hjellming} R.~M., {di
  Salvo} T., {van der Klis} M., 2004, \aap, 418, 255

\bibitem[{{Irwin} {et~al}\mbox{.}(1989){Irwin}, {Webster}, {Hewett},
  {Corrigan}, \& {Jedrzejewski}}]{1989ref89}
{Irwin} M.~J., {Webster} R.~L., {Hewett} P.~C., {Corrigan} R.~T.,
  {Jedrzejewski} R.~I., 1989, \aj, 98, 1989

\bibitem[{{Jackson}(2007)}]{2007ref7}
{Jackson} N., 2007, Living Reviews in Relativity, 10, 4

\bibitem[{{Jackson}(2011)}]{2011ref00}
{Jackson} N., 2011, \apjl, 739, L28

\bibitem[{{Jackson} {et~al}\mbox{.}(2000){Jackson}, {Xanthopoulos}, \&
  {Browne}}]{2000new43}
{Jackson} N., {Xanthopoulos} E., {Browne} I.~W.~A., 2000, \mnras, 311, 389

\bibitem[{{Kochanek}(1991)}]{1991new36}
{Kochanek} C.~S., 1991, \apj, 373, 354

\bibitem[{{Kochanek}(2006)}]{2006book2}
{Kochanek} C.~S., 2006, {Strong Gravitational Lensing}, pp. 91--+

\bibitem[{{Kochanek} {et~al}\mbox{.}(2006){Kochanek}, {Morgan}, {Falco},
  {McLeod}, {Winn}, {Dembicky}, \& {Ketzeback}}]{2006time34}
{Kochanek} C.~S., {Morgan} N.~D., {Falco} E.~E., {McLeod} B.~A., {Winn} J.~N.,
  {Dembicky} J., {Ketzeback} B., 2006, \apj, 640, 47

\bibitem[{{Koopmans} {et~al}\mbox{.}(2003){Koopmans}, {Biggs}, {Blandford},
  {Browne}, {Jackson}, {Mao}, {Wilkinson}, {de Bruyn}, \&
  {Wambsganss}}]{2003ref0}
{Koopmans} L.~V.~E. {et~al.}, 2003, \apj, 595, 712

\bibitem[{{Koopmans} {et~al}\mbox{.}(2009){Koopmans}, {Bolton}, {Treu},
  {Czoske}, {Auger}, {Barnab{\`e}}, {Vegetti}, {Gavazzi}, {Moustakas}, \&
  {Burles}}]{2009referans49}
{Koopmans} L.~V.~E. {et~al.}, 2009, \apjl, 703, L51

\bibitem[{{Koopmans} \& {de Bruyn}(2000)}]{2000ref86}
{Koopmans} L.~V.~E., {de Bruyn} A.~G., 2000, \aap, 358, 793

\bibitem[{{Koopmans} {et~al}\mbox{.}(2000){Koopmans}, {de Bruyn},
  {Xanthopoulos}, \& {Fassnacht}}]{2000time29}
{Koopmans} L.~V.~E., {de Bruyn} A.~G., {Xanthopoulos} E., {Fassnacht} C.~D.,
  2000, \aap, 356, 391

\bibitem[{{Koopmans} \& {Treu}(2002)}]{2002new37}
{Koopmans} L.~V.~E., {Treu} T., 2002, \apjl, 568, L5

\bibitem[{{Koopmans} {et~al}\mbox{.}(2006){Koopmans}, {Treu}, {Bolton},
  {Burles}, \& {Moustakas}}]{2006method17}
{Koopmans} L.~V.~E., {Treu} T., {Bolton} A.~S., {Burles} S., {Moustakas} L.~A.,
  2006, \apj, 649, 599

\bibitem[{{Marlow} {et~al}\mbox{.}(2001){Marlow}, {Rusin}, {Norbury},
  {Jackson}, {Browne}, {Wilkinson}, {Fassnacht}, {Myers}, {Koopmans},
  {Blandford}, {Pearson}, {Readhead}, \& {de Bruyn}}]{2001ref3}
{Marlow} D.~R. {et~al.}, 2001, \aj, 121, 619

\bibitem[{{Marshall} {et~al}\mbox{.}(2007){Marshall}, {Treu}, {Melbourne},
  {Gavazzi}, {Bundy}, {Ammons}, {Bolton}, {Burles}, {Larkin}, {Le Mignant},
  {Koo}, {Koopmans}, {Max}, {Moustakas}, {Steinbring}, \& {Wright}}]{2007new9}
{Marshall} P.~J. {et~al.}, 2007, \apj, 671, 1196

\bibitem[{{Myers} {et~al}\mbox{.}(1995){Myers}, {Fassnacht}, {Djorgovski},
  {Blandford}, {Matthews}, {Neugebauer}, {Pearson}, {Readhead}, {Smith},
  {Thompson}, {Womble}, {Browne}, {Wilkinson}, {Nair}, {Jackson}, {Snellen},
  {Miley}, {de Bruyn}, \& {Schilizzi}}]{1995ref58}
{Myers} S.~T. {et~al.}, 1995, \apjl, 447, L5

\bibitem[{{Myers} {et~al}\mbox{.}(2003){Myers}, {Jackson}, {Browne}, {de
  Bruyn}, {Pearson}, {Readhead}, {Wilkinson}, {Biggs}, {Blandford},
  {Fassnacht}, {Koopmans}, {Marlow}, {McKean}, {Norbury}, {Phillips}, {Rusin},
  {Shepherd}, \& {Sykes}}]{2003new21}
{Myers} S.~T. {et~al.}, 2003, \mnras, 341, 1

\bibitem[{{Myers} {et~al}\mbox{.}(1999){Myers}, {Rusin}, {Fassnacht},
  {Blandford}, {Pearson}, {Readhead}, {Jackson}, {Browne}, {Marlow},
  {Wilkinson}, {Koopmans}, \& {de Bruyn}}]{1999ref6}
{Myers} S.~T. {et~al.}, 1999, \aj, 117, 2565

\bibitem[{{Patnaik} {et~al}\mbox{.}(1992){Patnaik}, {Browne}, {Walsh},
  {Chaffee}, \& {Foltz}}]{1992ref7}
{Patnaik} A.~R., {Browne} I.~W.~A., {Walsh} D., {Chaffee} F.~H., {Foltz} C.~B.,
  1992, \mnras, 259, 1P

\bibitem[{{Pelt} {et~al}\mbox{.}(1994){Pelt}, {Hoff}, {Kayser}, {Refsdal}, \&
  {Schramm}}]{1994method7}
{Pelt} J., {Hoff} W., {Kayser} R., {Refsdal} S., {Schramm} T., 1994, \aap, 286,
  775

\bibitem[{{Pelt} {et~al}\mbox{.}(1996{\natexlab{a}}){Pelt}, {Kayser},
  {Refsdal}, \& {Schramm}}]{1996method9}
{Pelt} J., {Kayser} R., {Refsdal} S., {Schramm} T., 1996{\natexlab{a}}, \aap,
  305, 97

\bibitem[{{Pelt} {et~al}\mbox{.}(1996{\natexlab{b}}){Pelt}, {Kayser}, {Schild},
  \& {Thomson}}]{1996method8}
{Pelt} J., {Kayser} R., {Schild} R., {Thomson} D.~J., 1996{\natexlab{b}}, in
  IAU Symposium, Vol. 168, Examining the Big Bang and Diffuse Background
  Radiations, {M.~C.~Kafatos \& Y.~Kondo}, ed., p. 539

\bibitem[{{Rathna Kumar} {et~al}\mbox{.}(2013){Rathna Kumar}, {Tewes},
  {Stalin}, {Courbin}, {Asfandiyarov}, {Meylan}, {Eulaers}, {Prabhu}, {Magain},
  {Van Winckel}, \& {Ehgamberdiev}}]{2013ref79}
{Rathna Kumar} S. {et~al.}, 2013, ArXiv:1306.5105

\bibitem[{{Refsdal}(1964)}]{1964ref3}
{Refsdal} S., 1964, \mnras, 128, 307

\bibitem[{{Rusin} {et~al}\mbox{.}(2001){Rusin}, {Marlow}, {Norbury}, {Browne},
  {Jackson}, {Wilkinson}, {Fassnacht}, {Myers}, {Koopmans}, {Blandford},
  {Pearson}, {Readhead}, \& {de Bruyn}}]{2001ref8}
{Rusin} D. {et~al.}, 2001, \aj, 122, 591

\bibitem[{{Schneider} {et~al}\mbox{.}(1992){Schneider}, {Ehlers}, \&
  {Falco}}]{1992book1}
{Schneider} P., {Ehlers} J., {Falco} E.~E., 1992, {Gravitational Lenses}

\bibitem[{{Shepherd}(1997)}]{1997ref9}
{Shepherd} M.~C., 1997, in Astronomical Society of the Pacific Conference
  Series, Vol. 125, Astronomical Data Analysis Software and Systems VI,
  {G.~Hunt \& H.~Payne}, ed., p.~77

\bibitem[{{Snellen} {et~al}\mbox{.}(1995){Snellen}, {de Bruyn}, {Schilizzi},
  {Miley}, \& {Myers}}]{1995ref59}
{Snellen} I.~A.~G., {de Bruyn} A.~G., {Schilizzi} R.~T., {Miley} G.~K., {Myers}
  S.~T., 1995, \apjl, 447, L9

\bibitem[{{Suyu} {et~al}\mbox{.}(2013){Suyu}, {Auger}, {Hilbert}, {Marshall},
  {Tewes}, {Treu}, {Fassnacht}, {Koopmans}, {Sluse}, {Blandford}, {Courbin}, \&
  {Meylan}}]{2013ref77}
{Suyu} S.~H. {et~al.}, 2013, \apj, 766, 70

\bibitem[{{Suyu} {et~al}\mbox{.}(2010){Suyu}, {Marshall}, {Auger}, {Hilbert},
  {Blandford}, {Koopmans}, {Fassnacht}, \& {Treu}}]{2010time13}
{Suyu} S.~H., {Marshall} P.~J., {Auger} M.~W., {Hilbert} S., {Blandford} R.~D.,
  {Koopmans} L.~V.~E., {Fassnacht} C.~D., {Treu} T., 2010, \apj, 711, 201

\bibitem[{{Tewes} {et~al}\mbox{.}(2013){Tewes}, {Courbin}, {Meylan},
  {Kochanek}, {Eulaers}, {Cantale}, {Mosquera}, {Magain}, {Van Winckel},
  {Sluse}, {Cataldi}, {V{\"o}r{\"o}s}, \& {Dye}}]{2013ref76}
{Tewes} M. {et~al.}, 2013, \aap, 556, A22

\bibitem[{{van Haarlem} {et~al}\mbox{.}(2013){van Haarlem}, {Wise}, {Gunst},
  {Heald}, {McKean}, {Hessels}, {de Bruyn}, {Nijboer}, {Swinbank}, {Fallows},
  {Brentjens}, {Nelles}, {Beck}, {Falcke}, {Fender}, {H{\"o}randel},
  {Koopmans}, {Mann}, {Miley}, {R{\"o}ttgering}, {Stappers}, {Wijers},
  {Zaroubi}, {van den Akker}, {Alexov}, {Anderson}, {Anderson}, {van Ardenne},
  {Arts}, {Asgekar}, {Avruch}, {Batejat}, {B{\"a}hren}, {Bell}, {Bell}, {van
  Bemmel}, {Bennema}, {Bentum}, {Bernardi}, {Best}, {B{\^i}rzan}, {Bonafede},
  {Boonstra}, {Braun}, {Bregman}, {Breitling}, {van de Brink}, {Broderick},
  {Broekema}, {Brouw}, {Br{\"u}ggen}, {Butcher}, {van Cappellen}, {Ciardi},
  {Coenen}, {Conway}, {Coolen}, {Corstanje}, {Damstra}, {Davies}, {Deller},
  {Dettmar}, {van Diepen}, {Dijkstra}, {Donker}, {Doorduin}, {Dromer}, {Drost},
  {van Duin}, {Eisl{\"o}ffel}, {van Enst}, {Ferrari}, {Frieswijk}, {Gankema},
  {Garrett}, {de Gasperin}, {Gerbers}, {de Geus}, {Grie{\ss}meier}, {Grit},
  {Gruppen}, {Hamaker}, {Hassall}, {Hoeft}, {Holties}, {Horneffer}, {van der
  Horst}, {van Houwelingen}, {Huijgen}, {Iacobelli}, {Intema}, {Jackson},
  {Jelic}, {de Jong}, {Juette}, {Kant}, {Karastergiou}, {Koers}, {Kollen},
  {Kondratiev}, {Kooistra}, {Koopman}, {Koster}, {Kuniyoshi}, {Kramer},
  {Kuper}, {Lambropoulos}, {Law}, {van Leeuwen}, {Lemaitre}, {Loose}, {Maat},
  {Macario}, {Markoff}, {Masters}, {McFadden}, {McKay-Bukowski}, {Meijering},
  {Meulman}, {Mevius}, {Middelberg}, {Millenaar}, {Miller-Jones}, {Mohan},
  {Mol}, {Morawietz}, {Morganti}, {Mulcahy}, {Mulder}, {Munk}, {Nieuwenhuis},
  {van Nieuwpoort}, {Noordam}, {Norden}, {Noutsos}, {Offringa}, {Olofsson},
  {Omar}, {Orr{\'u}}, {Overeem}, {Paas}, {Pandey-Pommier}, {Pandey}, {Pizzo},
  {Polatidis}, {Rafferty}, {Rawlings}, {Reich}, {de Reijer}, {Reitsma},
  {Renting}, {Riemers}, {Rol}, {Romein}, {Roosjen}, {Ruiter}, {Scaife}, {van
  der Schaaf}, {Scheers}, {Schellart}, {Schoenmakers}, {Schoonderbeek},
  {Serylak}, {Shulevski}, {Sluman}, {Smirnov}, {Sobey}, {Spreeuw}, {Steinmetz},
  {Sterks}, {Stiepel}, {Stuurwold}, {Tagger}, {Tang}, {Tasse}, {Thomas},
  {Thoudam}, {Toribio}, {van der Tol}, {Usov}, {van Veelen}, {van der Veen},
  {ter Veen}, {Verbiest}, {Vermeulen}, {Vermaas}, {Vocks}, {Vogt}, {de Vos},
  {van der Wal}, {van Weeren}, {Weggemans}, {Weltevrede}, {White}, {Wijnholds},
  {Wilhelmsson}, {Wucknitz}, {Yatawatta}, {Zarka}, {Zensus}, \& {van
  Zwieten}}]{2013ref85}
{van Haarlem} M.~P. {et~al.}, 2013, \aap, 556, A2

\bibitem[{{Vuissoz} {et~al}\mbox{.}(2008){Vuissoz}, {Courbin}, {Sluse},
  {Meylan}, {Chantry}, {Eulaers}, {Morgan}, {Eyler}, {Kochanek}, {Coles},
  {Saha}, {Magain}, \& {Falco}}]{2008new24}
{Vuissoz} C. {et~al.}, 2008, \aap, 488, 481

\bibitem[{{Vuissoz} {et~al}\mbox{.}(2007){Vuissoz}, {Courbin}, {Sluse},
  {Meylan}, {Ibrahimov}, {Asfandiyarov}, {Stoops}, {Eigenbrod}, {Le Guillou},
  {van Winckel}, \& {Magain}}]{2007new23}
{Vuissoz} C. {et~al.}, 2007, \aap, 464, 845

\bibitem[{{Wambsganss} {et~al}\mbox{.}(1990){Wambsganss}, {Paczynski}, \&
  {Schneider}}]{1990ref91}
{Wambsganss} J., {Paczynski} B., {Schneider} P., 1990, \apjl, 358, L33

\bibitem[{{Xanthopoulos} {et~al}\mbox{.}(1998){Xanthopoulos}, {Browne}, {King},
  {Koopmans}, {Jackson}, {Marlow}, {Patnaik}, {Porcas}, \&
  {Wilkinson}}]{1998ref5}
{Xanthopoulos} E. {et~al.}, 1998, \mnras, 300, 649

\bibitem[{{York} {et~al}\mbox{.}(2005){York}, {Jackson}, {Browne}, {Koopmans},
  {McKean}, {Norbury}, {Biggs}, {Blandford}, {de Bruyn}, {Fassnacht}, {Myers},
  {Pearson}, {Phillips}, {Readhead}, {Rusin}, \& {Wilkinson}}]{2005ref2}
{York} T. {et~al.}, 2005, \mnras, 361, 259

\bibitem[{{Zhang} {et~al}\mbox{.}(1991){Zhang}, {Akujor}, {Chu}, {Mutel},
  {Spencer}, {Wilkinson}, {Alef}, {Matveyenko}, \& {Preuss}}]{1991new32}
{Zhang} F.~J. {et~al.}, 1991, \mnras, 250, 650

\end{thebibliography}

\end{document}